\documentclass[12pt]{article}
\usepackage{a4wide}
\usepackage{amssymb}
\usepackage{graphicx}
\usepackage{subfigure}
\usepackage{epsfig}
\usepackage[tbtags]{amsmath}
\usepackage{exscale}
\begin{document}
{\renewcommand{\thefootnote}{\fnsymbol{footnote}}
\hfill  IGC--11/5--1\\
\medskip
\begin{center}
{\LARGE  Black-hole horizons in modified space-time structures arising from canonical quantum gravity}\\
\vspace{1.5em}
Martin Bojowald,$^1$\footnote{e-mail address: {\tt bojowald@gravity.psu.edu}}
George M. Paily,$^1$\footnote{e-mail address: {\tt gmpaily@phys.psu.edu}}
Juan D. Reyes$^{2,3}$\footnote{e-mail address: {\tt jdreyes@matmor.unam.mx }}
and Rakesh Tibrewala$^4$\footnote{e-mail address: {\tt rtibs@imsc.res.in}}
\\
\vspace{1em}
$^1$Institute for Gravitation and the Cosmos,
The Pennsylvania State
University,\\
104 Davey Lab, University Park, PA 16802, USA\\
\vspace{1em}
$^2$Instituto de Matem\'aticas,\\
Universidad Nacional Aut\'onoma de M\'exico, Campus Morelia,\\
A. Postal 61-3, C.P. 58090, Morelia, Michoac\'an, Mexico\\
\vspace{0.5em}
$^3$Instituto de F\'isica y Matem\'aticas,\\
Universidad Michoacana de San Nicol\'as de Hidalgo,\\
Edificio C-3, Ciudad Universitaria, Morelia, Michoac\'an, Mexico\\
\vspace{1em}
$^4$The Institute of Mathematical Sciences,
CIT Campus, Chennai 600113, India
\end{center}
}

\setcounter{footnote}{0}

\newcommand{\case}[2]{{\textstyle \frac{#1}{#2}}}
\newcommand{\lP}{\ell_{\mathrm P}}

\newcommand{\md}{{\mathrm{d}}}
\newcommand{\tr}{\mathop{\mathrm{tr}}}
\newcommand{\sgn}{\mathop{\mathrm{sgn}}}

\newcommand*{\R}{{\mathbb R}}
\newcommand*{\N}{{\mathbb N}}
\newcommand*{\Z}{{\mathbb Z}}
\newcommand*{\Q}{{\mathbb Q}}
\newcommand*{\C}{{\mathbb C}}

\newcommand{\Ef}{E^\varphi}
\newcommand{\Kf}{K_\varphi}

\newcommand{\kp}{\ensuremath{K_\varphi}}
\newcommand{\kx}{\ensuremath{K_x}}
\newcommand{\ex}{\ensuremath{E^x}}
\newcommand{\ep}{\ensuremath{E^\varphi}}
\newcommand{\gp}{\ensuremath{\Gamma_\varphi}}
\newcommand{\mr}{\ensuremath{\sqrt{\frac{2M}{x}}}}
\newcommand{\mrl}{\ensuremath{\sqrt{2Mx}}}
\newcommand{\dkp}{\ensuremath{\,\delta\!K_\varphi}}
\newcommand{\dkx}{\ensuremath{\,\delta\!K_x}}
\newcommand{\dep}{\ensuremath{\,\delta\!E^\varphi}}
\newcommand{\dex}{\ensuremath{\,\delta\!E^x}}
\newcommand{\dn}{\ensuremath{\,\delta\!N}}
\newcommand{\dnx}{\ensuremath{\,\delta\!N^x}}
\newcommand{\dnt}{\ensuremath{\,\delta_2 N}}
\newcommand{\dgop}{\ensuremath{\delta_1 \Gamma_\varphi}}
\newcommand{\dgtp}{\ensuremath{\delta_2 \Gamma_\varphi}}
\newcommand{\dkop}{\ensuremath{\delta_1 K_\varphi}}
\newcommand{\dktp}{\ensuremath{\delta_2 K_\varphi}}
\newcommand{\dktx}{\ensuremath{\delta_2 K_x}}
\newcommand{\detp}{\ensuremath{\delta_2 E^\varphi}}
\newcommand{\detx}{\ensuremath{\delta_2 E^x}}
\newcommand{\dntx}{\ensuremath{\,\delta_2\!N^x}}
\newcommand{\be}{\begin{equation}}
\newcommand{\ee}{\end{equation}}
\newcommand{\dif}{\mathrm{d}}
\newcommand{\dkpp}{\ensuremath{\,\delta\!K'_\varphi}}
\newcommand{\dkxp}{\ensuremath{\,\delta\!K'_x}}
\newcommand{\depp}{\ensuremath{\,\delta\!E^{\varphi'}}}
\newcommand{\dexp}{\ensuremath{\,\delta\!E^{x'}}}
\newcommand{\dnxp}{\ensuremath{\,\delta\!N^{x'}}}
\newcommand{\dktpp}{\ensuremath{\delta_2 K'_\varphi}}
\newcommand{\dktxp}{\ensuremath{\delta_2 K'_x}}
\newcommand{\detpp}{\ensuremath{\delta_2 E^{\varphi '}}}
\newcommand{\detxp}{\ensuremath{\delta_2 E^{x'}}}

\begin{abstract}
  Several properties of canonical quantum gravity modify space-time
  structures, sometimes to the degree that no effective line elements
  exist to describe the geometry. An analysis of solutions, for
  instance in the context of black holes, then requires new insights.
  In this article, standard definitions of horizons in spherical
  symmetry are first reformulated canonically, and then evaluated for
  solutions of equations and constraints modified by inverse-triad
  corrections of loop quantum gravity.  When possible, a space-time
  analysis is performed which reveals a mass threshold for black holes
  and small changes to Hawking radiation.  For more general
  conclusions, canonical perturbation theory is developed to second
  order to include back-reaction from matter. The results 
  shed light on the questions of whether renormalization of
  Newton's constant or other modifications of horizon conditions
  should be taken into account in computations of black-hole entropy
  in loop quantum gravity.
\end{abstract}

\section{Introduction}

Canonical quantum gravity allows for new quantum space-time structures
to replace the classical continuum underlying general relativity.  If
new forms of space-time are realized, they must come along with a
modified version of general covariance embodied in the set of
transformations acting on them. Covariance in this general sense is
encoded in the constraint algebra of a gravitational theory under
consideration, which can easily acquire quantum corrections. Provided
the quantum constraints form a first-class algebra, covariance is not
broken but might be modified compared to the classical notion. Many of
the standard effects of space-time are then expected to change; this
article provides an exploration in the context of black-hole aspects.
(Qualitatively, there are similarities to deformed Poincar\'e
symmetries \cite{DSR}, but a detailed relationship is not
straightforward to work out.)

Loop quantum gravity \cite{Rov,ThomasRev,ALRev} has provided a
discrete notion of quantum space, subject to dynamical laws. Several
key properties of this kind of quantum geometry have implications that
can be implemented consistently at the level of modified classical
constraints, especially in the spherically symmetric context. The
general theory is not unique, and so several different types of
modifications are possible. But many characteristic features arise in
a way rather insensitive to quantization ambiguities, which are then
interesting to probe in concrete models. In this article, we focus on
corrections arising from the quantization of inverse components of the
triad variables used in loop quantum gravity. They imply corrections
in the Hamiltonian constraint \cite{QSDI} which change the dynamics
and, generically, the constraint algebra of effective geometries. In
this special case of spherical symmetry, also versions of corrected
constraints not changing the algebra exist, which provide interesting
comparisons.

We use these examples to probe possible influences of quantum
space-time on properties of non-rotating black holes. Concrete
scenarios of collapse or evaporation sensitive to global space-time
structure cannot be considered reliable at the present stage of
developments, but an exploration of general aspects is
worthwhile. Among them are: the form of solutions, the status of
different gauges, and direct implications for black holes such as
horizons, Hawking radiation, or singularities. Many of the classically
known properties can no longer be taken for granted when even the
notion of space-time has changed. The examples provided here thus show
some of the expectations toward quantum gravity in general, as well as
the way in which loop quantum gravity at present can deal with
them. 

In particular, we will see that classical horizon conditions require
modifications in order to produce dynamically consistent results in
the presence of quantum-gravity corrections. Section~\ref{s:Models}
introduces the corrections and discusses their consistency. In
Section~\ref{s:Undeformed} we find and analyze background solutions
analogous to the Schwarzschild and Painlev\'{e}--Gullstrand forms of
space-time. Section~\ref{s:SecondOrder} is devoted to second-order
perturbation theory in canonical form, directly applicable to
corrected constraints encoding modified space-time
structure. Section~\ref{s:Inverse}, finally, applies this perturbation
theory to horizon conditions with the main result: Classical horizon
conditions can consistently be used when constraints are corrected but
obey the classical algebra. When the constraint algebra is modified,
however, classical horizon conditions become inconsistent and
gauge-dependent. We will provide modifications of the horizon
conditions so as to make them consistent in the cases considered
here. The area-mass relation of horizons is corrected in both cases of
corrected constraints, with modified and unmodified constraint
algebra. In the final section we will discuss implications for
black-hole entropy calculations in loop quantum gravity, which so far
use the classical conditions after an implicit gauge fixing implied by
implementing the horizon as a boundary.

\section{Models in Connection Variables}
\label{s:Models}

We recall that the canonical set of variables used for a loop
quantization of gravity consists of the $\mathfrak{su}(2)$-valued
Ashtekar--Barbero connection $A_a^i(\bar{x})$ and a densitized triad
$E^a_i(\bar{x})$ (which can be seen as an $\mathfrak{su}(2)$-valued
densitized vector field) on a 3-dimensional manifold coordinatized by
$\bar{x}$.  (Here Latin indices $a,b,\ldots$ from the beginning of the
alphabet are space indices and those from the middle of the alphabet
$i,j,\ldots$ are internal indices.)

Classically, one gives a geometrical interpretation to these quantities
through their relation with the standard geometrical variables used in
the Hamiltonian formulation of general relativity: the Riemannian
metric $q_{ab}$ on space-like hypersurfaces embedded in spacetime with
normal $n^a$, and the corresponding extrinsic curvature
\be\label{ExCurv}
K_{ab}=\frac{1}{2}\mathcal{L}_n q_{ab}\,.
\ee
The spatial 3-dimensional metric $q_{ab}$
is constructed from the densitized triad via $(\det
q)q^{ab}=E_i^aE_i^b$, while the Ashtekar--Barbero connection
\cite{AshVar,AshVarReell} is related to the extrinsic curvature
 and the spin connection
$\Gamma^i_a$ compatible with the triad by the formula
\begin{equation} \label{AshtekarConnection}
A^i_a=\Gamma^i_a+\gamma K^i_a\,,
\end{equation}
where $K^i_a:=|\det E|^{-\frac{1}{2}}K_{ab}E^{bi}$ and $\gamma>0$ is
the Barbero-Immirzi parameter \cite{AshVarReell,Immirzi}. Its value,
estimated by calculations of black-hole entropy
\cite{ABCK:LoopEntro,Gamma}, is usually reported as $\gamma\sim 0.24$
\cite{Gamma2}. To that end, one treats the horizon as a boundary
  of space-time and imposes isolated-horizon conditions for the
  boundary fields. The boundary fields are then quantized and
  configurations giving rise to a certain value of the area are
  counted. Details of this procedure are still being developed
  \cite{KaulMaj,BHEntLog,BHEntRev,BHEntSU21,BHEntSU2I,BHEntSU2,BHEntSU2Rev} 
 but the basic premise of using classical horizon conditions to set up
 the quantum theory has not been questioned. Later in this article we
 will come back to the consistency issue arising from the fact that a
 classical condition for horizons is imposed before quantization.

We will simplify the dynamical discussion by working with a
midisuperspace model. Imposing spherical symmetry
\cite{SphKl1,SymmRed} and using adapted coordinates
$(t,x,\vartheta,\varphi)$ reduces the SU(2)-gauge of the original
variables to U(1). The densitized triad is then determined by two
U(1)-invariant functions $E^x(x)$ and $\Ef(x)$ and a pure gauge angle
$\eta(x)$. This gauge angle also determines the $x$-component of the
spin connection $\Gamma_x=-\eta'$, while its `angular' gauge-invariant
component is $\Gamma_\varphi=-E^x\,'/2\Ef$ (for details see
\cite{SphSymm,SphSymmHam}). Here and in what follows the prime denotes
a derivative with respect to the radial coordinate $x$, while a dot
will denote derivatives with respect to the time coordinate $t$.  The
spherically symmetric metric in terms of these variables is
\begin{equation} \label{SSMetric}
{\rm d}s^2=-N^2{\rm d}t^2+\frac{\Ef\,^2}{|E^x|}({\rm d}x+N^x{\rm
  d}t)^2+ |E^x|{\rm d}\Omega^2
\end{equation}
where $N(t,x)$ is the lapse function and $N^x(t,x)$ the only nonzero component of the shift vector.

Similarly the Ashtekar connection is determined by three functions: a
U(1)-connection $A_x(x)$, a U(1)-invariant function $A_\varphi(x)$
and the gauge angle $\eta(x)-\alpha(x)$. The relation (\ref{AshtekarConnection}) then gives \cite{SphSymmHam}
\begin{equation}
A_\varphi\cos\alpha=\gamma K_\varphi\quad,\quad
A_x+\eta'=\gamma K_x \quad\mbox{and}\quad
A_\varphi^2=\Gamma_\varphi^2+\gamma^2 K_\varphi^2 \notag
\end{equation}
for the gauge invariant parts $K_x$ and $K_\varphi$ of extrinsic curvature.
A suitable choice of  variables for a loop
quatization of spherically reduced gravity gives the symplectic structure:
\[
\{A_x(x),\frac{1}{2\gamma}E^x(y)\}=\{K_\varphi(x),E^\varphi(y)\}=\{\eta(x),\frac{1}{2\gamma}P^\eta(y)\}=G\delta(x,y)
\]
where  $P^{\eta}:=2A_{\varphi}E^{\varphi}\sin\alpha$ is the conjugate momentum of the gauge angle $\eta$.

Imposing the Gauss constraint
\[
G_{\rm grav}[\lambda]=\frac{1}{2G\gamma}\int \md x\,
\lambda((E^x)'+P^\eta)\,,
\]
the generator of the residual U(1)-gauge transformations of the
theory, we may further eliminate $\eta$ and $P^\eta$ and work with the
canonical pairs:
\be
\{\kx(x),\ex(y)\}=2G\delta(x,y)\quad\mbox{and}\quad\{\kp(x),\ep(y)\}=G\delta(x,y)\,.
\label{canonicalVars}
\ee

\subsection{Constraints and corrections} 
\label{constraints}

In these variables, the constraints of
general relativity (including generic matter contributions $\rho$ and
$J_x$ for the energy density and flux) take the following forms:
\begin{description}
\item[Hamiltonian Constraint]
\be\label{C2M}
\mathbf{H}[N]=-\frac{1}{2G}\int \md x\, N|\ex|^{-\frac{1}{2}}\bigg[\kp^2\ep+2\kp\kx\ex+(1-\gp^2)\ep+2\gp'\ex-8\pi G\ep|\ex|\rho \bigg]\approx0
\ee
\item[Diffeomorphism constraint]
\be\label{D2M}
\mathbf{D}[N^x]=\frac{1}{2G}\int \md x\,N^x\bigg[2\ep\kp'-\kx E^{x'}-8\pi G \ep \sqrt{|\ex|}J_x \bigg]\approx0
\ee
\end{description}

Under the conditions of the space-time being static, these constraints
and the equations of motion they generate can be solved to give the
traditional Schwarzschild metric, and similarly, the
Painlev\'e--Gullstrand metric, which describes the same situation in
different coordinates. We will demonstrate some of the derivations
below, including also one type of quantum corrections.  The
corrections discussed here are inspired by calculations in loop
quantum gravity, but we will attempt to keep our conclusions as
general as possible.  For the rest of this work we will use units
where the gravitational constant $G=1$.

Quantizing gravity is expected to lead to different types of quantum
corrections of the constraints. They are often subject to quantization
ambiguities, but requiring the new constraints to still have an
anomaly-free first class algebra puts restrictions on the general form
of the corrections. Anomaly-freedom, however, is not always easy to
achieve, and so it is useful to split an analysis of quantum-gravity
corrections into the different types. Even a single type of
correction, while not providing complete equations, can put strong
consistency conditions on the formalism. In loop quantum gravity, it
turns out that the easiest corrections to implement are the inverse
triad corrections, which arise when one quantizes terms in the
Hamiltonian constraint containing inverse components of the densitized
triad. In this framework, the $E^a_i$ are quantized to flux operators
with discrete spectra containing zero \cite{AreaVol,Area}. Since such
operators do not have densely defined inverses, no direct inverse
operator is available.  Instead well-defined techniques
\cite{QSDI,QSDV} are used that imply corrections (generally denoted as
$\alpha$ in what follows) to the classical inverse.

A basic condition on $\alpha$ is that it be a scalar to preserve the
spatial transformation properties of the corrected expressions. Among
the triad variables, $\ex$ is the only one with density weight zero,
and thus we can restrict $\alpha$ to depend only on $\ex$. In the
actual operators, $E^x(x)$ appears via fluxes, integrated over small
plaquettes forming the scaffolding of a discrete quantum state, not
via the whole orbit area $4\pi |E^x|$ at radius $x$. Qualitatively,
the correction function is a function $\alpha(\Delta)$ depending on
$E^x$ via the plaquette size $\Delta=E^x/{\cal N}(E^x)$, obtained by
dividing the orbit size by the number of plaquettes ${\cal N}(E^x)$
that form an orbit of size $|E^x|$. In general, the number must be
assumed to be a function of $E^x$ since a large orbit has to contain
more plaquettes than a smaller one in order to provide a similar
microscopic scale; this refinement of the underlying discrete
structure as the orbit considered grows is analogous to lattice
refinement in an expanding universe \cite{InhomLattice,CosConst}. As
in homogeneous models, lattice refinement cannot be fully derived
within a pure midisuperspace setting; the freedom must thus be
suitably parameterized.

\begin{figure}
\begin{center}
\includegraphics[width=12cm]{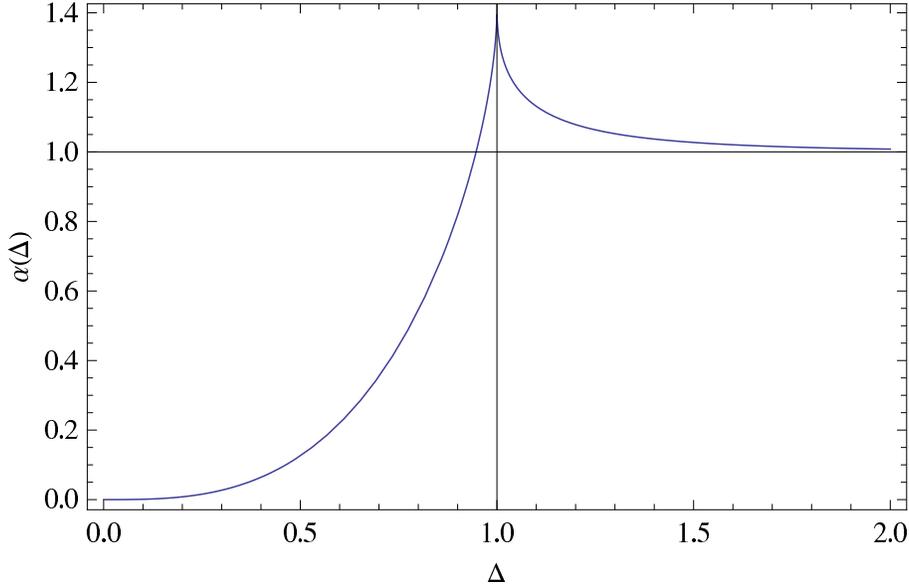}
\caption{\label{alphaDelta} The correction function $\alpha(\Delta)$ where $\Delta$ is taken relative to $\Delta_*:=\sqrt{\gamma{\cal N}/2}\lP$.}
\end{center}
\end{figure}

A concrete calculation of correction functions results in \cite{LTB}
\begin{equation} \label{alphaD}
\alpha(\Delta) = 2\sqrt{\Delta}\frac{\sqrt{|\Delta+\gamma \lP^2/2|}-
\sqrt{|\Delta-\gamma \lP^2/2|}}{\gamma \lP^2}
\end{equation}
with the Planck length $\lP$ (see Fig.~\ref{alphaDelta}).
(Quantization ambiguities imply that the functional form is not
uniquely fixed, but the qualitative properties used here are robust.)
For a small range of orbit radii and short evolution times, as
suitable for quasilocal horizon properties, one may assume ${\cal N}$
to be a constant, whose sole effect then is to raise the
quantum-gravity scale for $E^x$ where inverse-triad corrections become
important from $\lP^2$ to ${\cal N} \lP^2$.  These corrections
are thus relevant not just for Planck-size spheres; what matters
  is how close the size of elementary plaquettes is to the Planck
  scale.

In particular, including a gauge choice for $E^x$ we shall consider
the following situation:
\be \label{alpha without refinement}
\ex = x^2\quad, \quad \alpha = 2x\frac{\sqrt{|x^2+ \gamma{\cal N}
    \lP^2/2|}- \sqrt{|x^2-\gamma {\cal N}\lP^2/2|}}{\gamma {\cal
    N}\lP^2}\,.
\ee
As appropriate, we will comment throughout the paper on the
  reliability of our conclusions in light of the fact that not all
  possible quantum-gravity corrections are being considered.

The modified gravitational part of the Hamiltonian constraint we consider here is
\begin{align}
H^Q_{\rm grav}[N]=-\frac{1}{2G}\int \md x\, N\bigg[&\alpha|\ex|^{-\frac{1}{2}}\kp^2\ep+2\bar{\alpha}\kp\kx|\ex|^{\frac{1}{2}}+ \notag\\
&+\alpha|\ex|^{-\frac{1}{2}}(1-\gp^2)\ep+2\bar{\alpha}\gp'|\ex|^\frac{1}{2} \bigg] \label{myEffectiveHamiltonian}
\end{align}
with corrections of different powers of $E^x$ left independent by
using, for the sake of generality, two correction functions $\alpha$
and $\bar{\alpha}$. (Also $\gp$ might initially be expected to be
corrected with independent correction functions, but no such
anomaly-free version exists \cite{LTBII}.)  There is no inverse triad
component in the diffeomorphism constraint, and its action is directly
represented on graph states by the spatial deformations it generates.
Thus, we will keep the diffeomorphism constraint unmodified.

The  Poisson algebra of modified constraints closes:
\begin{eqnarray}
\{H_{\rm grav}^Q[N],D_{\rm grav}[N^x]\}&=&-H_{\rm grav}^Q[N^xN']\,,
\label{DefAgebraDiff}\\
\{H_{\rm grav}^Q[N],H_{\rm grav}^Q[M]\}&=&D_{\rm grav}[\bar{\alpha}^2|E^x|(\Ef)^{-2}(NM'-MN')]\,.   \label{DefAlgebra1}
\end{eqnarray}
The Poisson bracket relations (\ref{DefAlgebra1}) (or
(\ref{DefAlgebra2}) below in the presence of scalar matter) with
$\bar{\alpha}=1$ express the fact that dynamics takes place on
space-like hypersurfaces embedded in a pseudo-Riemannian spacetime
\cite{Regained}.  However, for generic $\bar{\alpha}\neq 1$, this
algebra no longer coincides with the algebra of spherically symmetric
hypersurface deformations of general relativity and generally
spacetime covariant systems.\footnote{Defining $\bar{\alpha}
N=\bar{N}$, we have
$\bar{\alpha}^{2}(NM'-MN')=(\bar{N}\bar{M}'-\bar{M}\bar{N}')$ and the
algebra can formally be written in classical form $\{H_{\rm
grav}^Q[\bar{N}/\bar{\alpha}],H_{\rm
grav}^Q[\bar{M}/\bar{\alpha}]\}=D_{\rm
grav}[|E^x|(\Ef)^{-2}(\bar{N}\bar{M}'-\bar{M}\bar{N}')]$. However,
using $\bar{N}$ in the full algebra will then modify the Poisson
bracket (\ref{DefAgebraDiff}) of the Hamiltonian with the
diffeomorphism constraint.}  The algebra is a fundamental object,
encoding not only gauge properties of gravity but the structure of
spacetime as well.  Thus, for $\bar{\alpha}\neq 1$ not only the
dynamics (reflected by the modified Hamiltonian constraint
(\ref{myEffectiveHamiltonian})) but also the structure or symmetries
of the spacetime manifold are changed by these quantum corrections.
Gauge symmetries will, in general, no longer coincide with coordinate
transformations in our models and we cannot interpret the dynamical
fields $\ex$ and $\ep$ as components of a pseudo-Riemannian metric as
in (\ref{SSMetric}): modified gauge transformations of $E^x$ and
$E^{\varphi}$ no longer match with coordinate transformations of ${\rm
d}x^{\mu}$ to form an invariant ${\rm d}s^2$. (Possible candidates for
space-time models corresponding to the corrected gauge transformations
are non-commutative manifolds or Finsler geometries.  The latter can
be explored in this context using the formalism of \cite{Finsler}.)

We will consider in detail the two cases $\alpha\not=\bar{\alpha}=1$
and $\alpha=\bar{\alpha}\not=1$. All other cases where
\be
 1\neq\alpha\neq\bar{\alpha}\neq 1
\ee
can formally be related to the solutions with $\bar{\alpha}=1$ by means
of the substitutions
\be
N\rightarrow \frac{N_{\alpha}}{\bar{\alpha}}; \quad \alpha \rightarrow \frac{\alpha}{\bar{\alpha}}.
\ee
where $N_{\alpha}$ refers to the value of $N$ in the $\bar{\alpha}= 1$
case. (We will keep matter terms general, referring only to the energy
density and energy flow without specific matter models. Thus, the
substitution will lead to different matter terms compared to the case
with $\bar{\alpha}=1$, but will not change the analysis of equations.)
Note, however, that since $\bar{\alpha}\neq 1$ in the original
constraints, the system is not generally covariant and coordinate
transformations between the different gauges do not exist.

For the first choice, the Hamiltonian constraint is
replaced by its modified counterpart (\ref{myEffectiveHamiltonian})
with $\bar{\alpha}=1$, while the diffeomorphism constraint (\ref{D2M})
remains unchanged, as does the overall form of the constraint
algebra. How does this first type of modification square with the
result \cite{Regained,LagrangianRegained} that given the constraint
algebra of classical general relativity, the Hamiltonian that depends
only on the 3-metric and the extrinsic curvature is uniquely the
classical Hamiltonian of general relativity? First note that
while the relations (\ref{canonicalVars}) still hold, we can no longer
interpret $K_a$  as extrinsic curvatures of the metric. This can be
easily seen from the equation for $\kx$ derived from the corrected evolution
equations, (\ref{EoMEx}) and (\ref{EoMEphi}) below.  Classically, we have:
\be
\kx=\frac{1}{N}\bigg(\frac{\dot{E}^{\varphi}}{\sqrt{|\ex|}}-\frac{\ep\dot{E}^x}{2|\ex|^{3/2}}+\frac{N^x\ep (\ex)'}{2|\ex|^{3/2}}-\frac{(N^x\ep)'}{\sqrt{|\ex|}}\bigg)
\ee
but with the modified equations we get: \be \label{Kx}
\kx=\frac{1}{N}\bigg(\frac{\dot{E}^{\varphi}}{\sqrt{|\ex|}}-\frac{\alpha\ep\dot{E}^x}{2|\ex|^{3/2}}+\frac{N^x\alpha\ep(\ex)'}{2|\ex|^{3/2}}-\frac{(N^x\ep)'}{\sqrt{|\ex|}}\bigg)
\ee which cannot be derived from relation (\ref{ExCurv}). The
component $\kp= (\dot{E}^x-N^x(E^x)')/2N\sqrt{|E^x|}$ is not modofied
if $\bar{\alpha}=1$.  We can relate the $\alpha$-modified $K_{a}$,
which are conjugate to the densitized triad, to the geometric
extrinsic curvature components from (\ref{ExCurv}), denoted here by
$\bar{K}_{a}$: \be \kp = \bar{K}_{\varphi} \quad,\quad \kx =
\bar{K}_x-\frac{(\alpha-1)\ep}{\ex} \bar{K}_{\varphi}.  \ee
Substituting this into (\ref{myEffectiveHamiltonian}), we get the
following Hamiltonian in terms of the densitized triad and the
extrinsic curvatures:
\begin{align}
H^Q_{\rm grav}[N]=-\frac{1}{2G}\int \md x\, N\bigg[&(2-\alpha)|
\ex|^{-\frac{1}{2}}\bar{K}_{\varphi}^2\ep+2\bar{K}_{\varphi}\bar{K}_x
|\ex|^{\frac{1}{2}}+ \notag\\
&+\alpha|\ex|^{-\frac{1}{2}}(1-\gp^2)\ep+2\gp'|\ex|^\frac{1}{2} \bigg] 
\end{align}
The Hamiltonian constraint is thus modified also in the geometrical
variables $\bar{K}_{a}$ which correspond to the classical form of
extrinsic curvature but are no longer canonically conjugate to the
densitized triads. Compared with the full situation
\cite{Regained,LagrangianRegained}, the constraint algebra in
spherical symmetry is thus not as restrictive, and different sets of
constraints can give rise to the same algebra.

\subsection{Matter fields}

For concreteness, we consider as a matter source a scalar field $\chi$
 with general potential $U(\chi)$. The matter part of the action we
 start with is
\[
S_{\rm matter}=-\frac{1}{2}\int {\rm d}^4x \sqrt{-\det g}\,[g^{\mu\nu}\partial_\mu\chi\partial_\nu\chi+U(\chi)]
\]
or in the 3+1 decomposition of the 4-metric  in terms of $q_{ab}$, lapse $N$ and shift vector $N^a$\begin{align}
S_{\rm matter}&=\int {\rm d}t\int
{\rm d}^3x\bigg[\Pi\dot{\chi}-N^a\Pi\partial_a\chi-N\left(\frac{\Pi^2}{2\sqrt{\det
      q}}+\frac{1}{2}\sqrt{\det
    q}q^{ab}\partial_a\chi\partial_b\chi+\frac{1}{2}\sqrt{\det q}U\right)\bigg] \notag
\end{align}
where $\Pi=\sqrt{\det q}(\dot{\chi}-N^a\partial_a\chi)/N$ is the canonically conjugate momentum of $\chi$.
{}From this form of the action we immediately identify the matter part of the diffeomorphism constraint and the kinetic, gradient and potential terms of the matter Hamiltonian.

Imposing spherical symmetry and defining $p_\chi$ by the relation $\Pi=p_\chi\sin \vartheta$ gives, after integration of the angular variables, the symplectic structure for a spherically symmetric scalar field minimally coupled to gravity:
\[
\{\chi(x),p_\chi(y)\}=\frac{1}{4\pi}\delta(x,y)\,.
\]
The matter contribution to the diffeomorphism constraint reads:
\[
D_{\rm matter}[N^x]=4\pi\int {\rm d}x\, N^x p_\chi \chi'
\]
and to the Hamiltonian constraint it is
\[
H_{\rm matter}[N]=\int {\rm d}x\,N(\tilde{\mathcal{H}}_\pi+\tilde{\mathcal{H}}_\nabla+\tilde{\mathcal{H}}_U)
\]
where the kinetic, gradient and potential terms are, respectively,
\[
\tilde{\mathcal{H}}_\pi=4\pi\,\frac{p_\chi^2}{2|E^x|^\frac{1}{2}\Ef} \quad,\quad
\tilde{\mathcal{H}}_\nabla=4\pi\,\frac{|E^x|^\frac{3}{2}\chi'\,^2}{2\Ef} \quad,\quad
\tilde{\mathcal{H}}_U=4\pi\,|E^x|^\frac{1}{2}\Ef\frac{U[\chi]}{2} \,.
\]

Since $\tilde{\mathcal{H}}_{\rm matter}=\sqrt{\det q}\,\rho$ and
$\tilde{D}_{\rm matter}=-\sqrt{\det q}\,J_x$, the energy density
$\rho=T_{ab}n^an^b$ and energy flux $J_a=q_a^bT_{bc}n^c$ are,
respectively,
\[
\rho=\frac{ p_\chi^2}{2|E^x|\Ef\,^2 }+\frac{|E^x|\chi'\,^2}{2\Ef\,^2}+\frac{U}{2}
\]
and
\[
J_x=-\frac{1}{|E^x|^\frac{1}{2}\Ef}p_\chi\chi'\,.
\]

Again, we introduce general quantum correction functions $\nu$
and $\sigma$ into the matter part of the Hamiltonian constraint
to account for the quantization of inverse-triad operators as:
\begin{equation} \label{HQmatter}
H_{\rm matter}^Q[N]=\int {\rm
  d}x\,N(\nu\tilde{\mathcal{H}}_\pi+\sigma\tilde{\mathcal{H}}_\nabla+
\tilde{\mathcal{H}}_U) \,.
\end{equation}
As before, only a dependence of the correction functions on $E^x$ is
possible for consistency with the unmodified diffeomorphism
constraint.  The potential term is not expected to acquire quantum
corrections because it does not contain an inverse of the triad. There
is no inverse of $E^x$ in the gradient term, either, which may thus be
expected to be unmodified by $E^{\varphi}$-independent
corrections. For generality, we nevertheless insert a second
correction function $\sigma(E^x)$ for this term (in contrast to the
potential term) because without spherical symmetry there is an
inverse-triad component in the gradient term and it would be
corrected. For all our subsequent calculations, it will nevertheless
be consistent to assume $\sigma=1$.

The presence of matter makes the constraint algebra more
non-trivial. In the gravitational part, one can sometimes absorb
correction functions in the lapse function if $\alpha=\bar{\alpha}$ at
least as far as the dynamics is concerned.  With a matter potential,
even if $\nu$ and $\sigma$ would equal $\alpha$, the correction does
not simply amount to a rescaling of the lapse function and the closure
of the constraint algebra becomes a nontrivial requirement that
restricts the form of the correction functions.  The total Hamiltonian
$\mathbf{H^Q}[N]=H_{\rm grav}^Q[N]+H_{\rm matter}[N]$ and
diffeomorphism constraint $\mathbf{D}[N^x]=D_{\rm grav}[N^x]+D_{\rm
matter}[N^x]$ satisfy the algebra:
\begin{eqnarray}
\{\mathbf{H^Q}[N],\mathbf{D}[N^x]\}&=&-\mathbf{H^Q}[N^xN']\,,\\
\{\mathbf{H^Q}[N],\mathbf{H^Q}[M]\}&=&D_{\rm grav}[\bar{\alpha}^2|E^x|(\Ef)^{-2}(NM'-MN')] \nonumber\\
&&+D_{\rm matter}[\nu\sigma|E^x|(\Ef)^{-2}(NM'-MN')]\,.  \label{DefAlgebra2}
\end{eqnarray}

The requirement of anomaly-freedom thus imposes the condition
\begin{equation} \label{anomalyFreeCond}
\bar{\alpha}^2=\nu\sigma
\end{equation}
and quantization ambiguities are somewhat reduced by relating
correction functions.

\subsubsection{Equations of Motion}

The canonical equations of motion obtained from the corrected
Hamiltonian are
\begin{align}
\dot{E}^x&=2N\bar{\alpha}\Kf|E^x|^\frac{1}{2}+N^x E^x\,' \label{EoMEx}\\
\dot{E}^\varphi&=N( \bar{\alpha}K_x|E^x|^\frac{1}{2}+\alpha\Kf\Ef|E^x|^{-\frac{1}{2}})+(N^x\Ef)' \label{EoMEphi} \\
\dot{\chi}&=\frac{N\nu}{|E^x|^\frac{1}{2}\Ef}p_\chi+N^x\chi'  \label{EoMchi} \\
\dot{p}_\chi&=\left(\frac{N\sigma|E^x|^\frac{3}{2}\chi'}{\Ef}\right)'-\frac{1}{2}N|E^x|^\frac{1}{2}\Ef\frac{\partial U}{\partial \chi}+(N^x p_\chi)'  \label{EoMpchi}
\end{align}

\begin{align}
\dot{K}_\varphi=&\frac{N}{2}|E^x|^{-\frac{1}{2}}\left[-\alpha\Kf^2+(2\bar{\alpha}-\alpha)\frac{E^x\,'\,^2}{4\Ef\,^2}-\alpha\right]+N^x\Kf'+(N\bar{\alpha})'\frac{|E^x|^\frac{1}{2}E^x\,'}{2\Ef\,^2} \notag\\
          &-2\pi G N\left[\nu\frac{p_\chi^2}{|E^x|^\frac{1}{2}\Ef\,^2}+\sigma\frac{|E^x|^\frac{3}{2}\chi'\,^2} {\Ef\,^2}-|E^x|^\frac{1}{2}U[\chi]\right]   \label{EoMKphi}
\end{align}

\begin{align}
\dot{K}_x=&-N\bar{\alpha}|E^x|^{-\frac{1}{2}}K_x\Kf+N\alpha\frac{|E^x|^{-\frac{3}{2}}\Ef}{2}\left(\Kf^2+1-\frac{E^x\,'\,^2}{4\Ef\,^2}\right) \notag\\
         &+N\bar{\alpha}|E^x|^{-\frac{1}{2}}\left(\frac{E^x\,''}{2\Ef}-\frac{E^x\,'\Ef\,'}{2\Ef\,^2}\right) +N(\bar{\alpha}-\alpha)\left(|E^x|^{-\frac{1}{2}}\frac{E^x\,'}{2\Ef}\right)' \notag\\
         &+\left[2(N\bar{\alpha})'-(N\alpha)'\right]\frac{|E^x|^{-\frac{1}{2}}E^x\,'}{2\Ef}-(N\bar{\alpha})'\frac{|E^x|^\frac{1}{2}\Ef\,'}{\Ef\,^2}+(N\bar{\alpha})''\frac{|E^x|^\frac{1}      		{2}}{\Ef} \notag\\
         &+(N^x K_x)'-N\frac{\partial \alpha}{\partial E^x}|E^x|^{-\frac{1}{2}}(\Kf^2\Ef+\Ef(1-\Gamma_\varphi^2))  \notag\\
         &-2N\frac{\partial \bar{\alpha}}{\partial E^x}|E^x|^\frac{1}{2}(K_x\Kf+\Gamma_\varphi')+2G N\left(\frac{\partial \nu}{\partial E^x}\tilde{\mathcal{H}}_\pi+\frac{\partial \sigma}			{\partial E^x}\tilde{\mathcal{H}}_\nabla\right) \notag\\
         &+2\pi G N\left(-\nu\frac{p_\chi^2}{|E^x|^\frac{3}{2}\Ef}+\sigma\frac{3|E^x|^\frac{1}{2}\chi'\,^2}{\Ef}+\frac{\Ef U[\chi]}{|E^x|^\frac{1}{2}}\right) \,.    \label{EoMKx}
\end{align}
We will later provide examples for their consistency, showing the
importance of conditions from anomaly freedom, including
(\ref{anomalyFreeCond}).

\section{Background solutions for undeformed constraint algebra}
\label{s:Undeformed}

For $\bar{\alpha}=1$ the constraints obey the classical algebra, and thus
generate coordinate changes as gauge transformations and allow the
existence of effective line elements to describe the modified
geometries. Anomaly-freedom then requires $\sigma=\nu^{-1}$, which
with the typical form (\ref{alphaD}) of inverse-triad corrections can
be satisfied only for $\sigma=1=\nu$. (Irrespective of quantization
ambiguities, inverse-triad correction functions have the
characteristic feature that they approach the classical value one from
above at large flux values \cite{Ambig,ICGC}; this cannot be satisfied
by both $\sigma$ and $\nu$ if they are mutual inverses and not equal
to one.) Thus, $\alpha$ is the only non-trivial correction function in
this case. In this subsection, we analyze its implications for
effective vacuum line elements. The usual properties of black holes
can then be studied by standard means; just corrections in metric
coefficients appear.

\subsection{Vacuum line elements}

For comparison and later reference, we derive vacuum solutions in two
commonly used spacetime gauges, producing line elements in the
Schwarzschild and Painlev\'e--Gullstrand form.

\subsubsection{Modified Schwarzschild metric}

In vacuum, we produce a Schwarzschild-type line element by imposing
the static gauge $K_x=K_{\varphi}=N^x=0$, and the diffeomorphism
constraint is automatically satisfied. The Hamiltonian constraint
requires
\begin{equation} \label{hamiltonian constraint in static gauge}
(1-\Gamma_{\varphi}^{2})\frac{\alpha
    E^{\varphi}}{\sqrt{E^{x}}}+2\Gamma'_{\varphi}\sqrt{E^{x}}=0\,.
\end{equation}
With the vanishing $K_{\varphi}$ obeying (\ref{EoMKphi}), we have the
further equation
\begin{equation} \label{derivative of lapse}
N'=\frac{N\alpha(E^{\varphi})^{2}}{E^{x}E^{x'}}+\frac{N\alpha
  E^{x'}}{4E^{x}}-\frac{NE^{x'}}{2E^{x}} \,.
\end{equation}
With these two differential equations for $E^{\varphi}$ and $N$ one
can check that (\ref{EoMKx}) is identically satisfied.

Next we specify the coordinate gauge $E^x=x^2$ so as to refer by $x$
to the area radius. Eq.~(\ref{hamiltonian constraint in static gauge})
then becomes
\begin{equation} \label{diff eq for ephi}
\alpha (E^{\varphi})^{3}-2x^{2}E^{\varphi}+2x^{3}E^{\varphi'}-\alpha
x^{2}E^{\varphi}=0 \,.
\end{equation}
With the classically motivated ansatz
$E^{\varphi}=x/\sqrt{1-2Mf_{\alpha}(x)/x}$ we
obtain the equation
\begin{equation} \label{diff eq for f}
\frac{f_{\alpha}'(x)}{f_{\alpha}(x)}=\frac{1-\alpha}{x} \,.
\end{equation}
The behavior of solutions to this equation for different refinement
schemes will be shown below. The functional form of the resulting line
element, which we first continue to derive, is largely independent of
the refinement scheme.

Using the solution for $E^{\varphi}$ along with the choice $E^{x}=x^{2}$ in \eqref{derivative of lapse} gives,
\begin{equation} \label{diff eq for n}
\frac{2N'x}{N}=\frac{\alpha}{1-2Mf_{\alpha}(x)/x}+\alpha-2\,.
\end{equation}
Again we use a classically motivated ansatz
$N=g_{\alpha}(x)\sqrt{1-2Mf_{\alpha}(x)/x}$ where $f_{\alpha}(x)$ is the
function found above, and obtain
\begin{equation} \label{diff eq for g}
\frac{g_{\alpha}'}{g_{\alpha}}=\frac{\alpha-1}{x}\,.
\end{equation}
Comparing this with \eqref{diff eq for f} we see that the solution for
$g_{\alpha}(x)$ is the inverse of the solution for $f_{\alpha}(x)$.
In what follows we will interchangeably use $g_{\alpha}(x)$ and
$1/f_{\alpha}(x)$.

We thus see that both $E^{\varphi}$ and $N$ pick up corrections due to
the inclusion of quantum effects. And since we already verified that
the condition $\dot{K}_{x}=0$ is satisfied assuming that the solution
is static, we have a valid solution for the modified Schwarzschild
line element:
\begin{equation} \label{Schw1}
{\rm d}s^{2}=-g_{\alpha}^{2}\left(1-\frac{2Mf_{\alpha}}{x}\right){\rm d}t^{2}+
\left(1-\frac{2Mf_{\alpha}}{x}\right)^{-1}{\rm d}x^{2}+x^{2}{\rm d}\Omega^{2}\,.
\end{equation}
Provided that $f_{\alpha}$ and $g_{\alpha}$ approach one in the asymptotic region of
large $x$, the classical Schwarzschild spacetime is recovered.  If one
were to use this solution all the way down to $x=0$, there is a
strongly modified region at small $x$, but the singularity at $x=0$
would not be resolved: The Ricci scalar, for one, diverges for the
usual functional behavior of $f_{\alpha}$. (Notice that the Ricci term
does not necessarily vanish even in vacuum if quantum corrections are
present.)

Between the asymptotic regime and the strongly modified one, we
encounter the possibility of horizon formation.  The equation for a
horizon is given by $2Mf_{\alpha}(x)=x$ or solving for $M$ we have that
$M=x/2f_{\alpha}(x)$ as the value of mass for which we have a horizon,
implicitly defined as a function of the horizon radius $x$.

We now look at the behavior of the solution for $f_{\alpha}$ for different
refinement schemes.

\paragraph{Constant patch number:}

First assuming ${\cal N}={\rm const}$, we solve Eq.~(\ref{diff eq for f}) 
for the two branches of $\alpha$ given by the absolute value in
(\ref{alpha without refinement}).  For $x^{2}>{\cal N}\gamma
\lP^{2}/2$,
\begin{equation} \label{sol for f when x is large}
f_{\alpha}(x)=\frac{2xe^{(1-\alpha)/2}}{\left(x+\sqrt{x^{2}-{\cal N}\gamma
    \lP^{2}/2}\right)^{1/2}\left(x+\sqrt{x^{2}+{\cal N}
\gamma \lP^{2}/2}\right)^{1/2}}
\end{equation}
where the constant of integration has been chosen by the requirement
that $\lim_{x\rightarrow\infty}f_{\alpha}(x)=1$. For $x^{2}<{\cal
N}\gamma l_{\rm Pl}^{2}/2$,
\begin{equation} \label{sol for f when x is small}
f_{\alpha}(x)=\frac{2e^{-\pi/4}x e^{(1-\alpha(x))/2}e^{\frac{1}{2}\arctan
 \left(\sqrt{x^{2}/({\cal N}\gamma\lP^{2}/2-x^{2})}\right)}}{(
 {\cal N}\gamma\lP^{2}/2)^{1/4}
(x+\sqrt{x^{2}+{\cal N}\gamma\lP^{2}/2})^{1/2}}
\end{equation}
where the constant of integration has been fixed by the requirement
that $f_{\alpha}(x)$ be continuous at $x^2={\cal N}\gamma \lP^{2}/2$.

\begin{figure}
\begin{center}
\includegraphics[width=12cm]{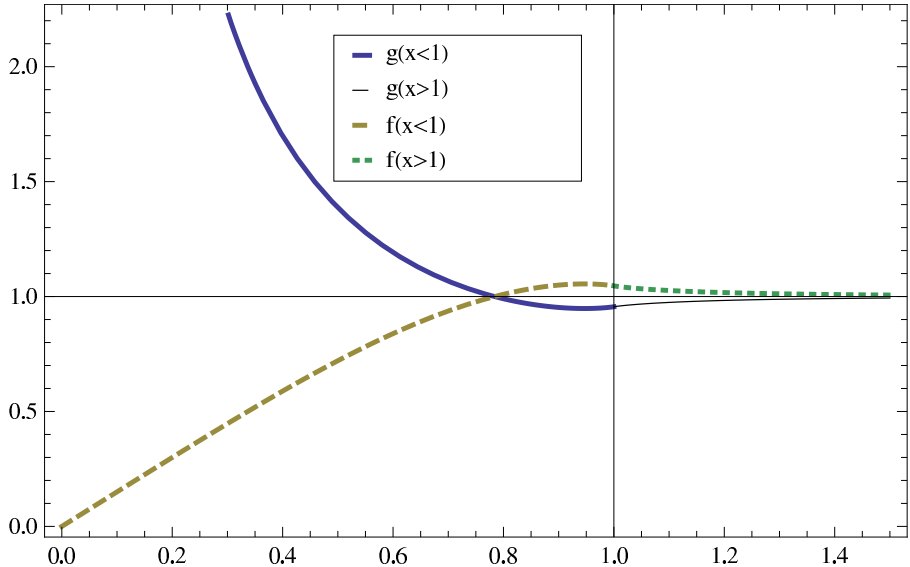}
\caption{\label{fgplot} Functions $g_{\alpha}(x)$ (solid) and
  $f_{\alpha}(x)$ (dashed) where $x$ is taken relative to
  $x_*:=\sqrt{\gamma{\cal N}/2}\ell_{\rm
  P}$.}
\end{center}
\end{figure}

\begin{figure}
\begin{center}
\includegraphics[width=12cm]{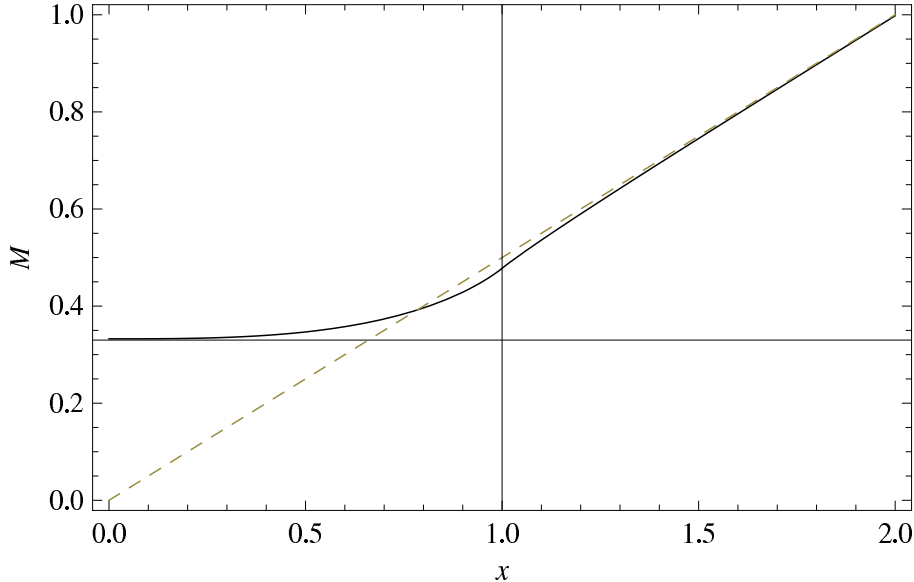}
\caption{\label{horizoncurve} Horizon curve: Right hand side of
  $M=x/2f_{\alpha}(x)$ (solid) and the classical horizon curve (dashed). The
  numerical values in this example are for ${\cal N}=1$; for larger ${\cal
  N}$ the mass threshold is raised by a factor of ${\cal N}^{1/4}$.}
\end{center}
\end{figure}

Figure \ref{fgplot} shows the behavior of $f_{\alpha}(x)$ and the
corresponding $g_{\alpha}(x)=1/f_{\alpha}(x)$ and we see that they
quickly tend to one beyond the scale $\sqrt{\gamma{\cal N}/2}\ell_{\rm
P}$, or the Planck length raised by the square root of the plaquette
number ${\cal N}$.  Figure \ref{horizoncurve} shows the graphical
solution of the horizon equation $M=x/2f_{\alpha}(x)$ for the example
${\cal N}=1$ and we see that there is a mass threshold below which no
horizon forms. For other (constant) values of ${\cal N}$ the size of
the mass threshold can easily be estimated as the limit
\begin{equation} \label{Mlim}
 \lim_{x\to 0} \frac{x}{2f_{\alpha}(x)}= \frac{1}{4} e^{\pi/4-1/2} ({\cal
   N}\gamma\lP^2/2)^{1/4}\,.
\end{equation}
Thus, for $M\lesssim \frac{1}{4} e^{\pi/4-1/2} {\cal N}^{1/4} \sim
0.33 {\cal N}^{1/4}$ (in Planck units with $\gamma$ absorbed) no
horizon forms. This observation agrees with results obtained
independently with quantum corrections of inverse-triad type
\cite{Collapse,HusainCritical,LTB,LTBII}. For comparison we have
also plotted the classical horizon curve and we see that the two
curves are nearly indistinguishable beyond $x=\sqrt{\gamma {\cal
    N}/2}\ell_{\rm P}$.

Given that small values of $x$ are associated with large curvature,
the solutions are probably no longer reliable all the way down to
$x=0$; other quantum corrections, ignored here, should be expected to
become strong as well. However, since $x/f_{\alpha}(x)$ is monotonic,
owing to ${\rm d}(x/f_{\alpha}(x))/{\rm d}x=
\alpha(x)/f_{\alpha}(x)\geq 0$, the limit (\ref{Mlim}) provides a
lower bound for the threshold to which the ratio $M=x/2f_{\alpha}(x)$
asymptotes. Since the modified curve starts to deviate from the
classical one at $x_*=\sqrt{\gamma{\cal N}/2}\lP$ which for large
${\cal N}$ need not be deep inside the quantum regime, the limit,
considered as an approximation for the asymptote value, gives a good
estimate for the mass threshold.

Since the small-$x$ regime is likely to require all corrections, in
addition to inverse-triad ones also quantum back-reaction and holonomy
corrections (which, too, can be argued to lead to a mass threshold
\cite{ModestoConformal}), an analysis as the present one cannot provide
hints for the full global structure or a conformal diagram.

\paragraph{Non-constant patch number:} If ${\cal N}$ is not constant
but depends on $x$, the correction functions $f_{\alpha}$ and
$g_{\alpha}$ change. With a power-law ansatz ${\cal N}\propto x^p$,
the qualitative behavior of importance here does not strongly depend
on $p$, except in the interesting case $p=2$ in which the patch size
(the orbit size divided by ${\cal N}$) is constant. We first present
the formulas for general $p=:2+\epsilon$, and then comment more
specifically on $p=2$ or $\epsilon=0$.

To be specific we choose a power-law behavior for the number of
plaquettes i.e. $\mathcal{N}= b^{2}x^{2+\epsilon}$. Here $b$ is a
constant with dimesnsions $[L^{-(1+\epsilon/2)}]$, introduced to ensure that
$\mathcal{N}$ is dimensionless. From \eqref{alpha without refinement}
we see that for the assumed form of $\mathcal{N}$ the correction
function $\alpha$ becomes
\begin{equation} \label{alpha with refinement}
\alpha=\frac{\sqrt{|1+a^{2}b^{2}x^{\epsilon}|}-\sqrt{|1-a^{2}b^{2}x^{\epsilon}|}}{a^{2}b^{2}x^{\epsilon}}\,,
\end{equation}
where we have introduced the notation $a^{2}\equiv \gamma
\ell_{\rm P}^{2}/2$. We note that the point separating the deep quantum
regime from the semiclassical regime is now dependent not just on the
Planck scale through $a$ but also on the constant $b$ and the exponent
$\epsilon$. 

It turns out that also for this form of $\alpha$ the
differential equation \eqref{diff eq for f} for $f_{\alpha}$ can be solved
exactly. The solution is
\begin{equation} \label{f with refinement one}
f_{\alpha}(x)=c_{1}e^{\alpha/\epsilon}\left(1+\sqrt{1-a^2b^2x^{\epsilon}}
\right)^{1/\epsilon}\left(1+\sqrt{1+a^2b^2x^{\epsilon}}
\right)^{1/\epsilon}
\end{equation}
for $1>a^{2}b^{2}x^{\epsilon}$ and
\begin{equation} \label{f with refinement two}
f_{\alpha}(x)=c_{2}x^{-1}\exp\left(\frac{\alpha}{\epsilon}+
\frac{1}{2\epsilon}\arctan\left(
\frac{-2+a^{2}b^{2}x^{\epsilon}}{2\sqrt{-1+a^{2}b^{2}x^{\epsilon}}}\right)
\right)\left(1+\sqrt{1+a^2b^2x^{\epsilon}} \right)^{1/\epsilon}
\end{equation}
for $1<a^{2}b^{2}x^{\epsilon}$. The constants $c_{1}$ and $c_{2}$ are
determined by imposing suitable boundary conditions. The condition
$1>a^{2}b^{2}x^{\epsilon}$, for which the solution in \eqref{f with
refinement one} has been written, would be valid for large $x$ only if
$\epsilon<0$. In this case $c_{1}$ is determined by requiring
$f_{\alpha}(x)\rightarrow 1$ in the limit $x\rightarrow\infty$, which gives
$c_{1}=(4e)^{-1/\epsilon}$. For $\epsilon=-2$ (and with $b=1$), which
is the case considered previously, this correctly gives
$c_{1}=2\sqrt{e}$ (see \eqref{sol for f when x is large}). The
constant $c_{2}$ is then determined by requiring continuity at
$x^{\epsilon}=1/a^{2}b^{2}$, which gives
$c_{2}=(4e)^{-1/\epsilon}(ab)^{1/\epsilon}e^{\pi/4\epsilon}$. For
$\epsilon=-2$, $c_{2}=2e^{1/2-\pi/8}/\sqrt{a}$.

It is easy to see that for $\epsilon>0$, the roles of the two
solutions are reversed and it is the second solution above which will
be valid for large values of $x$.  If we take the limit
$x\rightarrow\infty$ in \eqref{f with refinement two} with
$\epsilon>0$ we find that it diverges, implying that we do not have
the correct asymptotic limit. Another way to see this is to note that
in this case $\alpha_{x\to\infty}\rightarrow0$ whereas classically it
should approach one. Thus we see that the case with $\epsilon>0$ does
not correspond to physically acceptable solutions. Indeed, in this
case the patch size $\Delta=E^x/{\cal N}$ shrinks as one moves out to
larger radii, eventually falling into the regime where inverse-triad
corrections are large.

We now consider the case $\epsilon=0$ (i.e. $\mathcal{N} \propto
x^{2}$). In this case the correction function $\alpha$ turns out to be
a constant:
\begin{equation} \label{alpha with epsilon zero}
\alpha=\frac{\sqrt{1+a^{2}b^{2}}-\sqrt{|1-a^{2}b^{2}|}}{a^{2}b^{2}}\,.
\end{equation}
and the equation for $f_{\alpha}$ is solved by
\begin{equation} \label{f with epsilon zero}
f_{\alpha}(x)=c_{3}x^{1+a^{-2}b^{-2}\left(\sqrt{|1-a^{2}b^{2}|}-
\sqrt{1+a^{2}b^{2}}\right)}\,.
\end{equation}

In classical regimes, deviations of $\alpha$ from one should be small,
and we are led to choose $a^{2}b^{2}\ll 1$.  This corresponds to an
expression of $f_{\alpha}(x)$ proportional to $x$ raised to a very small
negative power. Thus, although the function diverges at $x=0$ and goes
to zero as $x\rightarrow\infty$, for a large range of radii it is very
nearly constant. Indeed, for $\mathcal{N} \propto x^{2}$ the size of
plaquettes on every orbit is the same. Since inverse-triad corrections
with refinement depend only on the plaquette size, they have the same
value for all orbits and do not drop off as $x\to\infty$. The only way
to make the corrections small in any finite range of $x$ is then by
choosing the proportionality constant $b^{2}$ in
$\mathcal{N}=b^{2}x^{2}$ to be small, which is implied by the choice
$a^{2}b^{2}\ll 1$.

Nevertheless, the asymptotic structure of the full space-time is
modified. First, one can check that the leading curvature invariants,
the Ricci scalar, the Ricci tensor squared and the Riemann tensor
squared, all vanish asymptotically. But asymptotic flatness is not
obviously realized in the given slicing: In (\ref{Schw1}), the
drop-off of $f_{\alpha}$ implies that the mass parameter $M$ plays no
role for large $x$, while the additional diverging factor of
$g_{\alpha}(x)=1/f_{\alpha}(x)$ becomes relevant. For large $x$, the
metric turns out to be conformally related to the flat metric: For
$g_{\alpha}(x)=g_0 x^{\delta}$ with a small, positive $\delta$, the
line element for large $x$ is
\begin{equation}
{\rm d}s^2\sim -g_{\alpha}^2{\rm d}t^2+{\rm d}x^2+x^2{\rm d}\Omega^2
 =y^{2\delta/(1-\delta)}\left(-g_0^2{\rm d}t^2+ (1-\delta)^{-2}{\rm
 d}y^2+y^2 {\rm d}\Omega^2\right)
\end{equation}
with $y:=x^{1-\delta}$. However, the conformal factor diverges at
$x\to\infty$. 

Thus, the simple-looking case of constant patch size (whose analog in
isotropic quantum space-times is often used in cosmological models)
implies non-trivial changes to the asymptotic form of the slicing
used. Even though quantum corrections do not become larger as the
asymptotic regime is approached, the cumulative effects of small
corrections over a large range of radii add up and may lead to
stronger effects in the asymptotic space-time. We leave a more
detailed analysis for future work as the present article is mainly
concerned with quasilocal horizon properties.

\subsubsection{Modified Painlev\'e--Gullstrand metric}
\label{s:ModPG}

With the classical constraint algebra satisfied for the corrections
with $\bar{\alpha}=1$, we can look for a coordinate transformation to
produce the Painlev\'e--Gullstrand form of the corrected metric
(\ref{Schw1}).  This coordinate system has as its time variable the
proper time measured by a freely falling observer in the Schwarzschild
spacetime (starting at rest from infinity and moving radially; see
e.g.\ \cite{PoissonToolkit}). To determine this proper time we proceed
as follows. The corrected Schwarzschild metric is independent of time
and therefore $\xi_{(t)}=\partial_{t}$ is a Killing vector. Now
consider the geodesic of a (radially) freely falling observer, with
the tangent to the geodesic denoted by $u^a$. Then we have
$u_a\xi^a_{(t)}=E$ constant. If we parameterize the
geodesic by its proper time $T$ and choose $E=-1$, we have
\begin{equation} \label{time component of geodesic}
g_{ab}u^a\xi^b_{(t)}=
-\frac{1}{f_{\alpha}^{2}}\left(1-\frac{2Mf_{\alpha}}{x}\right)
\frac{{\rm d}t}{{\rm d}T}=-1\,.
\end{equation}
In addition, $g_{ab}u^au^b=-1$, or
\begin{equation}
-\frac{1}{f_{\alpha}^{2}}\left(1-\frac{2Mf_{\alpha}}{x}\right)\left(\frac{{\rm d}t}{{\rm
 d}T}\right)^{2}+\left(1-\frac{2Mf_{\alpha}}{x}\right)^{-1}\left(\frac{{\rm
 d}x}{{\rm d}T}\right)^{2}=-1
\end{equation}
and with \eqref{time component of geodesic} we obtain
\begin{equation} \label{GeodesicX}
\frac{{\rm d}x}{{\rm d}T}=-\sqrt{f_{\alpha}^{2}-1+\frac{2Mf_{\alpha}}{x}}\,,
\end{equation}
where the negative sign for the square root corresponds to an
infalling observer. Thus, $u_a=(-1,
-(1-2Mf_{\alpha}/x)^{-1}\sqrt{f_{\alpha}^{2}-1+2Mf_{\alpha}/x},0,0)$ and
\begin{equation} \label{painleve time in schwarzschild coordinates}
{\rm d}T=-u_a{\rm d}x^a=
{\rm d}t+\left(1-\frac{2Mf_{\alpha}}{x}\right)^{-1}\sqrt{f_{\alpha}^{2}-1+
\frac{2Mf_{\alpha}}{x}}{\rm d}x \,.
\end{equation}
Inserting for ${\rm d}t$ from above in the Schwarzschild metric and
simplifying we arrive at the metric in Painleve-Gullstrand-like
coordinates
\begin{equation} \label{corrected metric in painleve coordinates}
{\rm d}s^{2}=-\frac{1}{f_{\alpha}^{2}}\left(1-\frac{2Mf_{\alpha}}{x}\right){\rm d}T^{2}+
f_{\alpha}^{-2}{\rm d}x^{2}+2f_{\alpha}^{-2}\sqrt{f_{\alpha}^{2}-1+\frac{2Mf_{\alpha}}{x}}{\rm d}T{\rm
  d}x+ x^{2}{\rm d}\Omega^{2}\,.
\end{equation}
(Notice that $T={\rm const}$-slices, which are classically flat, are
no longer so.)

In this derivation, we have made use of the fact that coordinate
changes are gauge transformations for this type of corrections, and
have used the usual geodesic properties in space-time. We can
explicitly verify the first property by checking that the constraints
are satisfied for the new form of the line element as well.  By
comparison with (\ref{SSMetric}) we obtain $E^{x}=x^{2}$ and
$E^{\varphi}=x/f_{\alpha}$, as well as $N=1$ and
$N^{x}=(f_{\alpha}^{2}-1+2Mf_{\alpha}/x)^{1/2}$. These when used in
\eqref{Kx} and the analogous equation for $K_{\varphi}$, obtained from
\eqref{EoMEx}, give
\begin{equation}
K_{\varphi} = -\sqrt{f_{\alpha}^{2}-1+\frac{2Mf_{\alpha}}{x}} \quad,\quad
K_{x} = \frac{\alpha f_{\alpha}+M\alpha/x-f_{\alpha}}{x\sqrt{f_{\alpha}^{2}-1+
2Mf_{\alpha}/x}}\,.
\end{equation}
The diffeomorphism constraint amounts to
$2E^{\varphi}K_{\varphi}'=K_{x}E^{x}{}'$ which is satisfied, as is the
Hamiltonian constraint. For later use, we note that $N=1$ appears to
be a suitable way to specify the Painlev\'e--Gullstrand form without
directly referring to space-time properties (while spatial flatness,
as seen, can be violated by quantum corrections).

\subsection{Space-time properties}

For the uncorrected algebra, the usual space-time notions can be used
to define and compute the position of horizons or other properties.
In this section we first calculate the surface gravity and then give a
detailed computation for the Hawking effect to show possible
implications of space-time modifications.

To calculate the surface gravity we
start by considering the 4-acceleration of a particle (of unit mass)
held stationary at radius $x$. For the case of a static, spherically
symmetric metric the only non-zero component of the acceleration is
$a^{x}=\Gamma^{x}_{tt}(u^{t})^{2}$ and the force required to hold the
particle at radius $x$ by a local agent is
$|a|=g_{xx}^{1/2}\Gamma^{x}_{tt}(u^{t})^{2}$.  The force required by
an agent at infinity, defining surface gravity, differs by a red-shift
factor of $g_{tt}^{1/2}$, $\kappa=a_{\infty}(x)=g_{tt}^{1/2}|a|$.  At
the horizon $x_{{\rm h}}=2Mf_{\alpha}$, and the surface gravity is
\begin{equation} \label{surface gravity at horizon}
\kappa|_{x_{{\rm h}}}=\left.\frac{\alpha}{4Mf_{\alpha}^{2}}\right|_{x_{{\rm h}}}\,.
\end{equation}
For later use, we note that it can also be written as
\begin{equation} \label{kappaC}
 \kappa= \frac{1}{2} \left.\frac{\partial C}{\partial x}\right|_{x=2Mf_{\alpha}}
\end{equation}
with $C=g_{\alpha}(1-2Mf_{\alpha}/x)$.

Since the solution is time independent we can go through the usual
derivation of the Hawking effect \cite{HawkingParticleCreation,
UnruhBlackHoleEvap, BirrellDavies}. We should note one crucial
difference expected for a complete picture: as the black hole
evaporates, its mass will decrease and will ultimately reach the
limiting value below which the horizon disappears giving a naked
singularity.  In this regime, however, we can no longer consider our
equations, or other corrections suggested by loop quantum gravity,
reliable. (See
\cite{HWCollapse,PolymerCollapse,StabilityLoopBH,EvapLoopBH} for other
calculations of horizons and evaporation with corrections motivated by
loop quantum gravity.)

We start by rewriting the metric as
\begin{equation}
{\rm d}s^{2}=g_{\alpha}^{2}\left(1-\frac{2Mf_{\alpha}}{x}\right)\left[-{\rm d}t^{2}+
  \frac{{\rm
  d}x^{2}}{g_{\alpha}^{2}\left(1-2Mf_{\alpha}/x\right)^2}\right]+x^{2}{\rm d}\Omega^{2}\,.
\end{equation}
Defining ${\rm d}x^{*}={\rm d}x/C={\rm d}x/g_{\alpha}\left(1-2Mf_{\alpha}/x\right)$
(with $C$ as above) and introducing the
null coordinates $u=t-x^{*}+X_{0}^{*}$, $v=t+x^{*}-X_{0}^{*}$
($X_{0}^{*}$ a constant), the metric becomes
\begin{equation} \label{modified metric in null coordinates}
{\rm d}s^{2}=\bar{C}(x){\rm d}u{\rm d}v
\end{equation}
where
$\bar{C}(x)=g_{\alpha}(x)C(x)=g_{\alpha}^{2}\left(1-2Mf_{\alpha}/x\right)$. We
now assume that this solution can be matched to a collapsing interior
given by the metric
\begin{equation} \label{interior}
{\rm d}s^{2}=A(U,V){\rm d}U{\rm d}V
\end{equation}
where $U=\tau-x+X_{0}$ and $V=\tau+x-X_{0}$ are the null coordinates
in the interior with $x=X_{0}$ being the surface of the star at $\tau
\leq 0$. In general, the surface of the star is given by
$x=X(\tau)$. To simplify the calculations, in what follows, we will
ignore the angular part of the metric and work in the 2-dimensional
$(t,x)$ space. As is usually done, we restrict ourselves to the $x
\geq 0$ region and impose the boundary condition that the scalar field
$\chi$ vanishes at $x=0$. In terms of the interior coodinates the line
$x=0$ is given by
\begin{equation} \label{x equal to zero line}
V=U-2X_{0}\,.
\end{equation}

We now solve the 2-dimensional massless scalar wave equation
$\Box\chi=0$ by functions that reduce to the standard form on
$\mathcal{I}^{-}$ and are subject to the boundary condition $\chi=0$
along \eqref{x equal to zero line}. If we let $U=a(u)$ and
$v=b(V)$ denote the identification of coordinates between the interior
and the exterior, then along $x=0$ we get
\begin{equation}
v=b(V)=b(U-2X_{0})=b(a(u)-2X_{0})\,. \nonumber
\end{equation}
This then gives the mode solution
\begin{equation} \label{in mode}
\chi_{\rm in}=i(4\pi \omega)^{-1/2}(e^{-i\omega
  v}-e^{-i\omega b[a(u)-2X_{0}]}) \,.
\end{equation}
The complicated $u$-dependence comes because the simple left-moving
wave is converted, due to the exponential redshift suffered by the
wave as the surface of the star approaches the horizon, to the
complicated out-going wave by the collapsing star. To determine the
free functions, we match the interior and the exterior metrics at the
boundary $x=X(\tau)$:
\[
g_{\alpha}C\left(-{\rm d}t^{2}+\left(\frac{{\rm d}x}{C}\right)^{2}\right) = A(-{\rm d}\tau^{2}+{\rm d}x^{2})
\]
such that
\[
\dot{t} = \frac{\left(ACg_{\alpha}^{-1}(1-\dot{X}^{2})+\dot{X}^{2}\right)^{1/2}}{C}
\]
where a dot here denotes derivative with respect to $\tau$. Since $U=a(u)$,
taking the derivative of the two sides we obtain ${\rm d}U = {\rm
d}\tau(1-\dot{X}) = a'{\rm d}u= a'(\dot{t}-\dot{X}/C){\rm d}\tau$, or
$a' = C(1-\dot{X})/(C\dot{t}-\dot{X})$. (In this section, a prime $(')$ denotes derivative with respect to the argument of the function.) Inserting $\dot{t}$,
\begin{equation} \label{alpha prime}
a'=\frac{C(1-\dot{X})}{\left(ACg_{\alpha}^{-1}(1-\dot{X}^{2})+\dot{X}^{2}\right)^{1/2}-\dot{X}}\,.
\end{equation}
The horizon is given by $C=0$ and therefore near the horizon, taking $(\dot{X}^{2})^{1/2}=-\dot{X}$ for a collapsing star, \eqref{alpha prime} simplifies to
\begin{equation} \label{alpha prime simple}
a'=\frac{C(\dot{X}-1)}{2\dot{X}}\,.
\end{equation}
(Since $g_{\alpha}^{-1}=f_{\alpha}$ is bounded from above, the validity of neglecting
the term involving $C$ is not affected by the presence of a factor of
$1/g_{\alpha}$.)  Also, near $C=0$ we expand $X(\tau)$ as $X(\tau)=X_{{\rm
h}}-\dot{X}(\tau_{{\rm h}})(\tau_{{\rm h}}-\tau)$ where the subscript
`h' designates the horizon. Using this we expand $C$ as
\[
C(X) = C(X_{{\rm h}}-\dot{X}(\tau_{{\rm h}})(\tau_{{\rm h}}-\tau))
= C(X_{{\rm h}})-\left.\frac{\partial C}{\partial x}\right|_{x_{{\rm h}}}\dot{X}(\tau_{{\rm h}})(\tau_{{\rm h}}-\tau)
= -\left.\frac{\partial C}{\partial x}\right|_{x_{{\rm h}}}\dot{X}(\tau_{{\rm h}})(\tau_{{\rm h}}-\tau)\,.
\]
Thus,
\begin{equation}
\frac{\partial{U}}{\partial{u}}=a'=-(\dot{X}-1)(\tau_{{\rm h}}-\tau)\kappa
\end{equation}
where $\kappa$ is given by (\ref{kappaC}).
We thus have
\begin{equation}
-\kappa {\rm d}u=\frac{{\rm d}U}{(\dot{X}-1)(\tau_{{\rm h}}-\tau)}\,.
\end{equation}

At the surface of the star, ${\rm d}U={\rm d}\tau(1-\dot{X})$ and
therefore we get
\begin{equation}
-\kappa u=\ln|\tau-\tau_{{\rm h}}|+c_{1}= \ln\vert U+X_{{\rm
 h}}-X_{0}-\tau_{{\rm h}}\vert+c_{1}
\end{equation}
where $c_{1}$ is a constant.
Near $\tau\approx\tau_{{\rm h}}$ we therefore obtain
\begin{equation} \label{alpha}
U =a(u)\propto e^{-\kappa u} + c_{2}
\end{equation}
where $c_{2}$ is some constant. We note that due to the negative
exponential, large changes in $u$ (the exterior coordinate at the
surface) near the horizon, where $u$ is large, correspond to small changes in $U$ (the
interior coordinate on the surface, near the horizon). Using similar
arguments the function $b$ relating $v$ and $V$, under the
assumption $C\approx0$, is found to be
\begin{equation} \label{beta prime}
b'=-\frac{A(1-\dot{X})}{2g_{\alpha}\dot{X}}\,.
\end{equation}
Due to the presence of a horizon, all the null rays corresponding to
$u= \rm constant$ near (and outside) the horizon, when traced back to
$\mathcal{I}^{-}$, correspond to a narrow range of $v= \rm constant$
rays. Similarly, because of the relation between $U$ and $V$ at the
surface of the star, a narrow range of $U$ values corresponds to a
narrow range in $V$ and therefore in the above equation one can assume
$A(U,V)$ to be constant. In this limit we also assume that
$g_{\alpha}$ is approximately a constant. (This assumption will not
always be justified since depending on the mass and the lattice
refinement scheme, the horizon could be at such a value of $x$ where
$g_{\alpha}$ could be a rapidly varying function. Here, the possibility of
stronger quantum effects arises which, however we will not explore in
this paper.) The equation can then be integrated easily to give
\begin{equation}
v=-\frac{A(1-\dot{X})}{2g_{\alpha}\dot{X}}V+c_{3}\,.
\end{equation}

Knowing the functions $a$ and $b$ we write the
complicated phase factor in \eqref{in mode} as
\begin{equation} \label{complex mode}
\chi_{\rm in}=i(4\pi\omega)^{-1/2}(e^{-i\omega v}-
e^{-i\omega[\bar{a}e^{-\kappa u}+\bar{b}]}) \,.
\end{equation}
Here $\bar{a}$ and $\bar{b}$ are some constants and as mentioned
above, $g_{\alpha}$ has been taken to be a constant and absorbed with these two
constants.

Instead of considering modes that are simple incoming waves on
$\mathcal{I}^{-}$ and complicated outgoing waves as in the equation
above, one may also consider modes which are simple outgoing waves
on $\mathcal{I}^{+}$ but which (when traced back to $\mathcal{I}^{-}$)
become complicated functions of $v$. To do so, one has to invert
the $u$-dependent phase factor in \eqref{complex mode}
above. It is straightforward to see that this gives the following
function of $v$
\begin{equation}
c(v)=-\frac{1}{\kappa}\ln\left\vert\frac{v-v_{0}}{\bar{a}}\right\vert
\end{equation}
where $v_{0}$ corresponds to the latest value of $v$ such that the
ray, starting on $\mathcal{I}^{-}$, reaches $\mathcal{I}^{+}$. This
also implies $v_{0}>v$. Thus this mode becomes
\begin{equation}
\chi_{\rm out}=i(4\pi\omega)^{-1/2}(e^{i\omega\kappa^{-1}\ln[(v_{0}-v)/\bar{a}]}-e^{-i\omega u}) \,.
\end{equation}

Knowing the `out' mode on $\mathcal{I}^{-}$, the next task is to
calculate the Bogolubov coefficients relating the two sets of modes
$\chi_{\rm in}$ and that portion of $\chi_{\rm out}$ that corresponds to waves
going to $\mathcal{I}^{+}$ for $v<v_{0}$, that is, $e^{-i\omega
c(v)}$. If the relation between the modes is given by
\begin{equation}
\chi_{\rm in}(\omega)=\int
{\rm d}\omega'\left(\alpha_{\omega\omega'}\chi_{\rm out}(\omega')+
\beta_{\omega\omega'}\chi^{*}_{\rm out}(\omega')\right) \,,
\end{equation}
then following the standard procedure, the relevant Bogolubov coefficient
describing particle production is $\beta_{\omega\omega'}$. In terms of
the standard inner product for scalar fields, this is given by
$\beta_{\omega\omega'}=-(\chi_{\rm in}(\omega),\chi_{\rm out}^{*}(\omega'))$. 
When evaluated this gives
\begin{equation} \label{spectrum}
\vert\beta_{\omega\omega'}\vert^{2}=\frac{1}{2\pi\kappa\omega}\left(\frac{1}{e^{\omega'/k_{\rm B}T-1}}\right)
\end{equation}
with $T=\hbar\kappa/2\pi k_{\rm B}$, $k_{\rm B}$ being the Boltzmann
constant. Using $\kappa$ as given by \eqref{kappaC} the temperature is
\begin{equation} \label{temperature}
k_{\rm B}T=\frac{\hbar\alpha(x_{{\rm h}}) g_{\alpha}^{2}(x_{{\rm h}})}{8\pi M}
\end{equation}
where the horizon is given by $C=1-2Mf_{\alpha}/x_{\rm h}=0$. In the limit
$x\rightarrow\infty$ (for large $M$) we recover the well known result
$k_{\rm B}T=\frac{\hbar}{8\pi M}$.

The classical formulas are thus corrected by several factors from
inverse-triad corrections. But there was also an additional
position-dependence in the derivation, which, in regimes where it
must be taken into account, makes the analysis more complicated but
might introduce new and stronger effects.

\section{Second-order perturbations}
\label{s:SecondOrder}

We now perform perturbative calculations for the classical vacuum
constraints, to be used in the context of matter back-reaction.
Although the classical vacuum constraints can be solved exactly, the
perturbative procedure as well as some of the equations will be useful
later. We will also take this opportunity to state our background gauge
conditions for the two versions of the space-time, Schwarzschild and
Painlev\'e--Gullstrand.  For each of the backgrounds considered,
we will perform the following steps:

\begin{description}
\item[Step 1] 1st-order perturbation of the Hamiltonian and diffeomorphism constraints.
\item[Step 2] 1st-order perturbation of the equations of motion.
\item[Step 3] 2nd-order perturbation of constraints including matter fields.
\item[Step 4] 2nd-order perturbation of the equations of motion, as necessary.
\item[Step 5] Calculation of the perturbed form of the metric and
  evaluation of horizon conditions to find area-mass relationships.
\end{description}

In addition, the following features are common to all the calculations:

\begin{itemize}
\item The perturbations of the fundamental variables ($\ex,\ep,\kx,\kp$) will be denoted as:
\be
\ep \rightarrow \ep + \Delta\ep = \ep + \dep + \detp
\ee
and so on. In every case, a $\delta$ without a subscript is to be
taken to refer to a first-order perturbation of the relevant
quantity. Fields without any kind of delta refer to the
  background values. 
\item Since we are interested in possible changes to the area of various surfaces, we will, for simplicity of calculation, make the gauge choice
    \be \label{Exgauge}
    \ex = x^2
    \ee
to fix the diffeomorphism constraint.
In particular, we set $\Delta\ex=0$, and the perturbation of $\ex$ at
    every order is set to zero. We will not be fixing the gauge
    completely. Rather, the presence of gauge-dependent terms
    (under transformations generated by the perturbed Hamiltonian
    constraint) will be 
    taken as one of the criteria to distinguish between the horizon
    conditions used in various models with different types of
    inverse-triad corrections. A key consistency requirement will be
    that horizon conditions be gauge invariant.
\item We consider the matter field and its corresponding conjugate
momentum to be first order perturbations (the background space-time is
vacuum), which implies that the energy density $\rho$ and the
energy-momentum flux $J^x$ are to be included only in the second order
and higher perturbations of the constraints.
\end{itemize}

The different slicings (Schwarzschild and Painlev\'e--Gullstrand) are
  implemented by specifying the background fields.  We will carry out
  the steps of the calculation in detail for the
  Painlev\'e--Gullstrand metric for an uncharged non-rotating black
  hole. For subsequent calculations we will only list the relevant
  changes. First, we provide two canonical versions of horizon
  conditions to be used.

\subsection{Horizon conditions}

We define horizons in canonical variables in order to be able to apply
them to equations corrected by effects from canonical quantum gravity.
For comparison, we provide two versions which would classically be
equivalent in the context of spherically symmetric geometries.  In
doing so, we must use space-time notions to capture the meaning of a
horizon, and it is not guaranteed that such definitions are reasonable
for models with a deformed constraint algebra and their new versions
of space-time structures. The motivation for providing two versions of
horizon conditions is that we can test whether they remain equivalent
in the deformed context and then have a chance of capturing the same
effects. For cases with an uncorrected constraint algebra, we will
furthermore compare with the direct space-time analysis.

\subsubsection{Trapping horizon}

Horizons of our perturbative solutions can be analyzed by an expansion
of the usual conditions, for instance of \cite{trapping}. In spherical
symmetry, the cross-section of a spatial slice with a spherical
trapping horizon as the boundary of spherical marginally trapped
surfaces, can be defined simply as a sphere at radius $x$ whose
co-normal $\dif x_a$ is null. This condition may be written as $g^{xx}=0$; one
can verify that zero expansion of null geodesics is then implied. In
triad variables with line element (\ref{SSMetric})
one obtains the condition
\be
\frac{\ex}{(E^\varphi)^2}-\left(\frac{N^x}{N}\right)^2=0
\ee
which can easily be analyzed perturbatively.
To second order in the perturbations, it expands to:
\be\label{horcond}
\begin{split}
&\bigg[\frac{\ex}{(E^{\varphi})^2}-\left(\frac{N^x}{N}\right)^2\bigg]_{0}\\
&+\bigg[-2\frac{\ex\dep}{(\ep)^3} - 2\frac{N^x \dnx}{N^2} + 2\frac{(N^x)^2 \delta\!N}{N^3}\bigg]_{1}\\
&+\bigg[-2\frac{\ex\detp}{(\ep)^3}-2\frac{N^x\dntx}{N^2}+2\frac{(N^x)^2 \delta_2 \!N}{N^3} + 3\frac{\ex (\dep)^2}{(\ep)^4} \\
&-\frac{(\dnx)^2}{N^2}-3\frac{(N^x \delta\!N)^2 }{N^4}+4\frac{N^x\dnx\delta\!N}{N^3}\bigg]_{2}=0\,.
\end{split}
\ee

\subsubsection{Isolated horizon}

Alternatively, for the Schwarzschild slicing we can define a spherical
horizon by using the specialization of isolated horizon conditions
\cite{HorRev} to spherical symmetry. Since matter is still allowed
outside the horizon, a situation comparable to the previous definition
is obtained, but the condition is more restrictive because no matter
is allowed at the horizon.

We are now dealing with the condition \cite{Horizon}
\be
A_{\varphi} = \sqrt{\kp^2 + \gp^2} = 0\,.
\ee
In the Schwarzschild metric this gives us two conditions:
\be\label{isoHorcond1}
\kp^2 = 0 \quad\mbox{and thus}\quad \dkp^2 = 0
\ee
and
\be\label{isoHorcond2}
\gp^2 = 0 \quad \mbox{or}\quad \gp^2 + 2\gp\delta\gp + 2\gp\dgtp +(\delta\gp)^2 = 0\,.
\ee
The fact that we have two conditions instead of just one as in
(\ref{horcond}) demonstrates the more restrictive notion. In spherical
symmetry, it turns out that the difference does not matter much
classically, but it will become important with spacetime-deforming
quantum corrections.

\subsubsection{Comparison and gauge}

The origin of the additional condition arising for isolated horizons
can be seen in the fact that isolated horizons, defined as boundaries
of space-time, freeze gauge transformations generated by the
Hamiltonian constraint on the horizon by boundary conditions. The
additional condition on $K_{\varphi}$ then formally replaces a possible
gauge-fixing condition one might choose in a treatment where the
horizon is not a boundary. Classically, the trapping-horizon condition
(\ref{horcond}) is gauge invariant, and its evaluation does not depend
on which gauge fixing is used. It thus implies results equivalent to
those produced by the isolated-horizon condition. 

However, it turns out that the condition (\ref{horcond}) is no longer
gauge invariant for some versions of quantum corrected
constraints. The horizon condition itself will then have to be
corrected so as to cancel the gauge dependence, thereby shedding some
light on what quantum horizon conditions could look like. For an isolated
horizon, on the other hand, having the Hamiltonian gauge fixed by
boundary conditions eliminates the important option of seeing how
horizon conditions must be corrected in addition to the dynamics of
quantum gravity. We will address these questions in detail by the
examples provided in the rest of this article.

\subsection{Painlev\'e--Gullstrand}

The Painlev\'e--Gullstrand form of the Schwarzschild space-time is
\begin{equation} \label{PG}
 {\rm d}s^2= -\left(1-\frac{2M}{x}\right) {\rm d}t^2+ {\rm d}x^2+
 2\sqrt{\frac{2M}{x}} {\rm d}t{\rm d}x+ x^2{\rm d}\Omega^2\,.
\end{equation}
It is characterized by several interesting properties, such as having
flat spatial slices of constant $t$.  In what follows, the background
solutions will appear as coefficients of perturbation equations,
partially identifying the gauge in which perturbations are analyzed.
For the Painlev\'e--Gullstrand background,
\[
\Ef=x, \quad N=1, \quad N^x=\sqrt{\frac{2M}{x}}, \quad K_x= \sqrt{\frac{M}{2x^3}}, \quad \text{and} \quad \Kf=-\sqrt{\frac{2M}{x}}
\]
in addition to (\ref{Exgauge}).

\subsubsection{First order perturbation of the constraints} \label{FirstOrderPerturbationSection1}

We expand the Hamiltonian and diffeomorphism constraint equations
$\delta \mathbf{H}[N]/\delta N=0$ and $\delta
\mathbf{D}[N^x]/\delta N^x=0$ to first order in metric
perturbations, obtaining the general forms
\begin{align}
2(&\kp\ep +\kx\ex)\dkp +2\kp\ex\dkx + (\kp^2-\Gamma_{\varphi}^2+1)\dep \notag\\
 &+ 2(\kp\kx+\gp')\dex   -2\gp \ep \delta\gp +2\ex\delta\gp' =0 \label{C1}
\end{align}
and
\be\label{D1}
2\ep\dkp'+2\kp'\dep-\kx\dexp-E^{x'}\dkx=0
\ee

Inserting the unperturbed form of the densitized triad and extrinsic
curvature corresponding to (\ref{PG}), and applying the gauge condition $\dex = 0$, with the additional corollaries that $\dexp=0$ and $\delta\Gamma_\varphi=\frac{E^x\,'}{2\Ef\,^2}\delta\Ef$, we have:
\be \label{CPG}
 -\mrl\dkp - 2\sqrt{2Mx^3}\dkx +\frac{2M}{x}\dep  +2x\depp = 0
\ee
and
\be \label{DPG}
2\mrl\dkpp+\frac{2M}{x^2}\dep-2\mrl\dkx=0\,.
\ee

To proceed solving the equations as far as possible, we subtract
(\ref{CPG}) and (\ref{DPG}) to obtain
\begin{equation}
\depp=\frac{1}{2}\mr\dkp+\mrl\dkpp=(\sqrt{2Mx}\,\dkp)'\,,
\end{equation}
which can immediately be integrated. If we impose the boundary conditions that all the perturbations fall off to zero at infinity, and in particular, that
\be \label{KphiFalloffCondition}
\sqrt{x}\dkp\rightarrow 0 \quad\mbox{as}\quad x\rightarrow\infty\, ,
\ee
this equation can be solved by
\begin{equation} \label{EphiPG}
\dep=\mrl\dkp \,,
\end{equation}
and, substituting this back in (\ref{DPG})
\be
\dkx=\frac{M}{x^2}\dkp+\dkpp\,.
\ee

\subsubsection{Perturbation of the Equations of motion} \label{FirstOrderPerturbationSection2}

We obtain the linear equations of motion by expanding the general
spherically symmetric equations  (\ref{EoMEx})--(\ref{EoMKx})
with $\bar{\alpha}=\alpha=1$.
Equation (\ref{EoMEx}) gives to first order
\be
\delta\dot{E}^x=2|\ex|^\frac{1}{2}(\kp\delta\!N+N\dkp)+
N\kp|\ex|^{-\frac{1}{2}}\dex+N^x\dexp+E^{x'}\dnx
\ee
or
\be
\delta\dot{E}^x=-2x\mr\delta\!N+2x\dkp-\frac{1}{x}\mr \dex+\mr\dexp+2x\dnx
\ee
with the background
solution (\ref{PG}) inserted for the unperturbed variables.  We have
$\delta\dot{E}^{\varphi}=\sqrt{2Mx} \delta\dot{K}_{\varphi}$ from
(\ref{EphiPG}). Using the equations of motion, this provides a second
relation between $\delta N$, $\delta K_{\varphi}$ and $\delta N^x$
which turns out to be identically satisfied.

To implement the gauge for the perturbations, we set $\dex$ and all its derivatives to zero,
to give:
\be
-\mr\dn+\dkp+\dnx=0\,.
\ee

If we make the further choice that $\dn=0$, we can use the
relations derived above to arrive at simplified equations for the
other perturbations; in particular $\delta N^x=-\delta\Kf$ and:
\be \label{dKphiPDE}
\dot{\dkp}=\mr\dkpp-\frac{1}{2x}\mr\dkp\,.
\ee
The first order set of equations is solved by the general solution to
(\ref{dKphiPDE}):
\[
\delta\Kf=\sqrt{x}\,F\left(2x^{3/2}/3+\sqrt{2M}\,t\right)
\]
for an arbitrary function $F$ of one variable as indicated, satisfying
the asymptotic condition (\ref{KphiFalloffCondition}). However, this
extra gauge condition $\dn=0$ is not necessary for our later results.
 The expressions for \dkx, \dep\, and \dnx\, in terms of
\dkp\, are consistent, and satisfy equations (\ref{EoMKphi}) and
(\ref{EoMKx}) for $\dot{\dkp}$ and $\dot{\dkx}$.

\subsubsection{Second order perturbation of the constraints including matter}

The second-order diffeomorphism constraint including matter is
\be \label{d2DiffConstraintGeneral}
2\delta\Ef\delta\Kf'+2\Ef \delta_2\Kf' +2 \Kf'\delta_2\Ef-E^x\,'\delta_2 K_x-8\pi\Ef|E^x|^\frac{1}{2}J_x=0\,,
\ee
so in our coordinates and using first order results we have
\be \label{pdiffeo2}
4M\dkp\dkpp+2\sqrt{2Mx}\,\dktpp+\frac{2M}{x^2}\detp-2\sqrt{2Mx}\,\dktx - 8\pi x^2\sqrt{\frac{2M}{x}} J_x =0\, .
\ee

The second-order Hamiltonian constraint
\begin{align}
(&\Kf^2-\Gamma_\varphi^2+1)\delta_2\Ef+2\Kf E^x\delta_2 K_x +2(\Kf\Ef+K_xE^x)\delta_2\Kf \notag \\
&-2\Gamma_\varphi\Ef\delta_2\Gamma_\varphi+2E^x\delta_2\Gamma_\varphi'
+2\Kf\delta\Kf\delta\Ef+\Ef(\delta\Kf)^2 \notag \\
+&2E^x\delta\Kf\delta K_x-2\Gamma_\varphi\delta\Gamma_\varphi\delta\Ef-\Ef(\delta\Gamma_\varphi)^2-8\pi\Ef|E^x|\rho=0  \label{d2HConstraintGeneral}
\end{align}
with $\delta_2\Gamma_\varphi=\frac{E^x\,'}{2\Ef\,^2}\delta_2\Ef-\frac{E^x\,'}{2\Ef\,^3}(\delta\Ef)^2$, requires a little more work and gives
\be
\begin{split} \label{SOHConstraint}
&-\mr\dktp-\sqrt{8Mx}\dktx+\frac{2M}{x^2}\detp+2\detpp+
\bigg((x-4M)(\dkp)^2\bigg)' -8\pi x^2\rho =0\, ,
\end{split}
\ee

Subtracting these constraints,
\be
2\detpp=\mr\dktp+2\sqrt{2Mx}\,\dktpp-((x-6M)(\dkp)^2)' + 8\pi x^2 \mathcal{H}
\ee
and integrating gives
\be \label{delta2Ephi0}
\detp=\mrl\dktp-\frac{1}{2}(x-6M)(\dkp)^2-\frac{1}{2}\int_x^\infty \mathrm{d}z\: 8\pi z^2  \mathcal{H}
\ee
where we have used
\be
\mathcal{H}:=N\rho-N^{x}J_{x}\, .
\ee

\subsubsection{Second order perturbation of the equations of motion}

We may proceed putting (\ref{delta2Ephi0}) back into the
diffeomorphism constraint (\ref{pdiffeo2}) to get an equation for
$K_x$ in terms of $\Kf$.  Equation (\ref{EoMEx}) gives
 \be
 -\sqrt{\frac{2M}{x}}\delta_2N+\delta_2\Kf+\delta_2N^x+\delta N\delta\Kf=0
 \ee 
and (\ref{EoMEphi}), upon using these and the first order equations, results in an evolution
equation for $\Kf$ consistent with equation (\ref{EoMKphi}). Since we
will not use these equations for the horizon conditions we will not
write them here.

\subsubsection{Perturbation of the metric and horizon}

After inserting the relevant expressions into (\ref{horcond}), we find
that the zeroth 
order terms are naturally the same as for the background, the first
order terms vanish --- 
which is to be expected since the matter terms have not yet played a
part --- and the second order terms include an influence from the
matter fields. The condition on the horizon becomes:
\be
1-\frac{2M}{x} +\frac{2}{x}\int_x^\infty \mathrm{d}z\:4\pi z^2 \mathcal{H}  = 0
\ee
which tells us that
\be
R_{\rm hor} = 2\bigg(M-\int_{R_{\rm hor}}^\infty \mathrm{d}z\:4\pi z^2 \mathcal{H}\:  \bigg)\,.
\ee
The horizon radius is simply shifted from the vacuum value $2M$ in
terms of the asymptotic mass by the amount of energy contributed by
matter between the horizon and spatial infinity. The dependence on
$\Delta K_{\varphi}$ in some solutions, for instance in
(\ref{delta2Ephi0}), automatically cancels when they are combined to
the horizon condition: the resulting condition is gauge invariant.

\subsection{Schwarzschild}
\label{s:S}

We proceed with the calculations in the Schwarzschild metric in a manner analogous to the Painlev\'e--Gullstrand case.

\subsubsection{Step 1}

In the Schwarzschild metric, assuming $\dex=0$, the first order Hamiltonian constraint (\ref{C1}) can be simplified to
\be
\bigg( 2 \bigg(1-\frac{2M}{x}\bigg)^{3/2}\dep \bigg)'=0\,.
\ee
The simplest solution to satisfy this constraint is to have
$\dep =
c(1-2M/x)^{-3/2}$. However this
choice blows up near the horizon faster than
$\ep=x(1-2M/x)^{-1/2}$, so we
make the choice $\dep=0$.

In the Schwarzschild gauge, the first order diffeomorphism constraint (\ref{D1}) becomes
\be \label{SDiffCon1}
\dkpp = \sqrt{1-\frac{2M}{x}} \dkx \,.
\ee
This relation will be used repeatedly to simplify the second order
constraints and equations of motion.

\subsubsection{Step 2}

>From the first order perturbation of equation (\ref{EoMEx}) for $\ex$ we derive
\be \label{d1NxS}
\dnx=-\sqrt{1-\frac{2M}{x}}\dkp\,.
\ee
Considering (\ref{EoMEphi}), the equation of motion for $\ep$,  we find:
\be
\dot{\dep}=\dkp + x\sqrt{1-\frac{2M}{x}}\dkx + \bigg( x
  \bigg(1-\frac{2M}{x} \bigg)^{-1/2}\dnx \bigg)'  \notag
\ee
which, using (\ref{SDiffCon1}) and (\ref{d1NxS}),  simplifies to
\be
\dot{\dep}=\dkp+x\dkpp+(-x\dkp)' =0
\ee
and ensures that $\dep$ remains zero.

Finally, equation (\ref{EoMKphi}) gives the additional relation
\be
\delta\dot{\Kf}=\left(1-\frac{2M}{x}\right)\delta N'-\frac{M}{x^2}\delta N
\ee

\subsubsection{Step 3}

The second order Hamiltonian constraint (\ref{d2HConstraintGeneral}), after
simplification and discarding terms containing $\dep$, becomes
\be
  \bigg(2 \bigg(1-\frac{2M}{x} \bigg)^{3/2} \detp \bigg)' + \bigg(
  x(\dkp)^2 \bigg)'- 8\pi x^2\rho =0
\ee
and implies
\be
2\bigg(1-\frac{2M}{x} \bigg)^{3/2}\detp + x(\dkp)^2 = -\int_{x}^{\infty}\dif z\:{8\pi z^2\rho  }\,.
\ee

The relation provided by the diffeomorphism constraint (\ref{d2DiffConstraintGeneral}) and the second order equations of motion are not needed here to derive the horizon condition, so we may proceed directly to step 5.  

\subsubsection{Step 5}
The condition on the horizon is:
\be
1-\frac{2M}{x} +\frac{2}{x}\int_x^\infty \mathrm{d}z\:4\pi z^2 \mathcal{H}  = 0
\ee
where in the Schwarzschild slicing $\mathcal{H}=\rho$. This agrees with our Painlev\'e--Gullstrand result.

Additionally, we can use the isolated horizon conditions, and we find that (\ref{isoHorcond2}) gives, after setting $\dep = 0$:
\be \label{schwisohorcondeq1}
1-\frac{2M}{x} + \dkp^2
-\frac{2}{x}\bigg(1-\frac{2M}{x}\bigg)^{3/2}\detp = 0\,.
\ee
But since $\dkp^2 = 0$ from (\ref{isoHorcond1}) at the isolated horizon, 
we once again have:
\be\label{schwisohorcondeq2}
1-\frac{2M}{x} + \frac{2}{x}\int_x^{\infty}{\dif z\, 4\pi z^2\rho } =0\,.
\ee
showing that we get equivalent results for the two methods of deriving the position of the horizon.

\section{Inverse-triad corrections}
\label{s:Inverse}

We are especially interested in horizon conditions in the presence of
back-reaction and quantum corrections. For $\bar{\alpha}=1$ the
constraints satisfy the classical hypersurface-deformation algebra
despite the presence of corrections. Effective line elements can thus
be used to describe the space-time geometry and standard horizon
definitions are applicable. We will first evaluate these definitions
in the presence of corrections, which still provide equivalent
results. This outcome is non-trivial since the modified dynamics could
have led to stronger changes of the horizon behavior, rendering different
definitions inequivalent. Moreover, the results of horizon conditions
will be gauge invariant.

For $\alpha=\bar{\alpha}\not=1$ we have a modified constraint algebra
but can obtain horizon formulas simply by substitution after absorbing
$\alpha$ in the lapse function as far as the gravitational part of the
Hamiltonian constraint is concerned.  (There are still non-trivial
quantum corrections: Matter Hamiltonians are non-classical even if we
absorb $\alpha$ in the lapse function for the gravitational part,
unless $U=0$ and $\nu=\sigma=\alpha$ in (\ref{HQmatter}).)

The most interesting case is thus that of
$1\not=\bar{\alpha}\not=\alpha$, which as stated previously can be
related to these two special cases. Here, the standard horizon
conditions will no longer be gauge invariant, but we present a
modification leading to satisfactory results. We will come back to
conclusions drawn from this case in the discussions.

\subsection{Classical algebra}
\label{s:Class}

Modified dynamics in the presence of ordinary space-time structure can
directly be evaluated by the canonical horizon definitions.

\subsubsection{Modified Painlev\'e--Gullstrand gauge}

We consider the modified Painlev\'e--Gullstrand metric (\ref{corrected metric in painleve coordinates}) as our background. The correction
function $\alpha$ depends only on $\ex$, so by assuming $\dex=0$, we
also have $\delta \alpha =0$. We will use the short hand
notation:
\be
h: = f_{\alpha}^2 -1 + \frac{2Mf_{\alpha}}{x}\,.
\ee

\paragraph{Step 1}
\subparagraph{Modified Hamiltonian constraint H[N]:}
To first order, assuming $\dex=0$, the modified Hamiltonian constraint reads
\begin{equation}  \label{quantumH1}
\begin{split}
 2(\alpha\kp\ep + &\kx\ex)\dkp+2\kp\ex\dkx+ \alpha( K_\varphi^2 -\Gamma_{\varphi}^2+1)\dep \\
 &  -2\alpha\gp \ep  \delta\gp +2\ex\delta\gp'=0\,.
\end{split}
\end{equation}
For the modified Painlev\'e--Gullstrand metric, using the relation between $\alpha$ and $f_{\alpha}$, this simplifies to:
\be
\bigg(\frac{x\sqrt{h}}{f_{\alpha}}\bigg)' \dkp + x\sqrt{h}\dkx -\frac{f_{\alpha}^2}{x}(x\dep)' +\frac{f_{\alpha}}{x}\bigg(\frac{x}{f_{\alpha}}\bigg)'\bigg(\frac{-M f_{\alpha}}{x}+f_{\alpha}^2\bigg)\dep =0\,.
\ee

\subparagraph{Diffeomorphism constraint $D[N^x]$:}

Equation (\ref{D1}) becomes:
\be
\frac{x\sqrt{h}}{f_{\alpha}}\dkpp -\sqrt{h}(\sqrt{h})'\dep - x\sqrt{h}\dkx =0\,.
\ee
Adding these first order equations, we get an expression that simplifies to:
\be
\bigg(\frac{x\sqrt{h}}{f_{\alpha}}\dkp\bigg)' - (f_{\alpha}^2\dep)' = 0
\ee
which implies, with the appropriate fall off conditions at infinity, that
\be
\dep=\frac{x\sqrt{h}}{f_{\alpha}^3}\dkp\,.
\ee

\paragraph{Step 2}

The equation of motion (\ref{EoMEx}) for $\ex$ gives us the relation
\be
-\sqrt{h}\dn + \dkp + \dnx = 0
\ee
in this modified Painlev\'e--Gullstrand metric.

\paragraph{Step 4}

Similarly for the second order perturbation of the
same equation, we derive
\be
\dn\dkp - \sqrt{h}\dnt + \dktp + \delta_2 N^x = 0\,.
\ee

\paragraph{Step 3}

Adding the second order constraints, integrating and rearranging, we get
\be
\frac{2x\sqrt{h}}{f_{\alpha}}\dktp -
\bigg(\frac{x}{f_{\alpha}}-\frac{3xh}{f_{\alpha}^3}\bigg)(\dkp)^2-2f_{\alpha}^2\detp
= \int_{x}^{\infty}\dif z\:  8\pi z^2 \frac{\mathcal{H}}{f_{\alpha}}
\ee
where
\be
 \mathcal{H}=N{\rho}-N^x J_x\,.
\ee

\paragraph{Step 5}
The condition on the horizon in the modified metric is:
\be \label{HorizonfPG}
1-\frac{2Mf_{\alpha}}{x} + \frac{2f_{\alpha}}{x}\int_x^{\infty}{\dif z\: \,4\pi z^2
  \frac{\mathcal{H}}{f_{\alpha}} } =0
\ee
which agrees with the classical Painlev\'e--Gullstrand result in the
limit that $f_{\alpha} \rightarrow 1$. Gauge-dependent terms such
  as $\delta K_{\varphi}$ drop out and there is no need to fix the
  Hamiltonian gauge.

\subsubsection{Modified Schwarzschild gauge}

\paragraph{Step 1}

\subparagraph{Modified Hamiltonian constraint $H[N]$:}

Using the relation between $\alpha$ and
$f_{\alpha}$, equation (\ref{quantumH1}) simplifies to:
\be
\bigg( \frac{2}{f_{\alpha}} \bigg(1-\frac{2Mf_{\alpha}}{x}\bigg)^{3/2}
  \dep \bigg)'=0 \,.
\ee
The simplest solution to satisfy this constraint is to have $\dep =0$.

\subparagraph{Diffeomorphism constraint $D[N^x]$:}

Equation (\ref{D1}) for this metric gives the relation:
\be
\dkpp = \sqrt{1-\frac{2Mf_{\alpha}}{x}} \dkx\,.
\ee

\paragraph{Step 2}

{}From the first order perturbation of the equation of motion for
$\ex$, we have:
\be
\dnx=-\frac{\dkp}{f_{\alpha}}\sqrt{1-\frac{2Mf_{\alpha}}{x}}\,.
\ee
For the equation of motion of $\dep$, we find
\be
\dot{\dep}=
\bigg(-\frac{x}{f_{\alpha}}\dkp\bigg)'+\frac{x}{f_{\alpha}}(\dkp)'+\bigg(\frac{1}{f_{\alpha}}-\frac{x
  f_{\alpha}'}{f_{\alpha}^2}\bigg)\dkp =0
\ee
which, once again, ensures that $\dep$ remains zero.

\paragraph{Step 3}

The second order Hamiltonian constraint, after simplification and
discarding terms which contain $\dep$, becomes:
\be \label{HamConsf}
 \bigg(\frac{2}{f_{\alpha}}\bigg(1-\frac{2Mf_{\alpha}}{x}\bigg)^{3/2}\detp 
 \bigg)' + \bigg( \frac{x(\dkp)^2}{f_{\alpha}} \bigg)'- \frac{8\pi x^2\rho}{f_{\alpha}} =0\,,
\ee

As in the classical Schwarzschild case, the relations from the second
order diffeomorphism constraint (\ref{d2DiffConstraintGeneral}) and
equations of motion are not needed to derive the expression for the
horizon condition.

For the horizon condition, we arrive at
\be \label{HorizonfS}
1-\frac{2Mf_{\alpha}}{x} +\frac{2f_{\alpha}}{x}\int_x^{\infty}{\dif z
  \;4\pi z^2\frac{\rho}{f_{\alpha}}} =0
\ee
which agrees with the classical Schwarzschild result in the limit that
$f_{\alpha}\rightarrow 1$. Equivalent results, (\ref{HorizonfPG}) and
(\ref{HorizonfS}) are obtained with both slicings and, in the
Schwarzschild slicing, with both definitions of horizons. Moreover,
for vacuum the result is in agreement with the direct space-time
analysis performed in Sec.~\ref{s:Undeformed}, which applies in this
subsection where the classical algebra is assumed in the presence of
corrections. In both cases, $\delta K_{\varphi}$-terms automatically
cancel in the horizon equation.

\subsection{Modified algebra, absorbable}
\label{s:ModAbs}

Before we evaluate horizon conditions in the case of a modified
constraint algebra, we present calculations that show the overall
consistency of the equations of motion and constraints.  We will
perform some of the calculations explicitly for the choice $N\alpha=1$
with a scalar matter field, illustrating how the anomaly-freedom
condition is necessary to obtain consistent equations. (See the
  Appendix for an illustration of the inconsistency of line elements
  in this case with modified space-time structures.)

\subsubsection{Dynamical consistency}

First-order equations and results for this case are identical to those
in sections \ref{FirstOrderPerturbationSection1} and
\ref{FirstOrderPerturbationSection2}. The second-order diffeomorphism
constraint is the same as (\ref{pdiffeo2}), and in the second order
Hamiltonian constraint (\ref{SOHConstraint}) the matter term is
replaced by
$-2x\alpha^{-1}(\nu\tilde{\mathcal{H}}_\pi+\sigma\tilde{\mathcal{H}}_\nabla+\tilde{\mathcal{H}}_U)$.
Again, combining these equations and integrating gives 
\begin{equation} \label{delta2Ephi}
\delta_2\Ef=\sqrt{2Mx}\,\delta_2\Kf-\frac{1}{2}(x-6M)(\delta\Kf)^2-\mathcal{E}
\end{equation}
where now we use the short hand notation
\begin{align}
\mathcal{E}:=&\,\int_x^\infty {\rm d}z\,4\pi z^2(N\rho_{\rm mod}\,-N^xJ_x) \notag\\
=&\,4\pi \int_x^\infty {\rm d}z\left[\frac{1}{\alpha}\left(\frac{\nu}{2z^2}p_\chi^2+\sigma\frac{z^2}{2}\chi'\,^2+\frac{z^2}{2}U\right)+\sqrt{\frac{2M}{z}}p_\chi\chi'\right]\,. \label{effectiveE}
\end{align}
Putting (\ref{delta2Ephi}) back into the diffeomorphism constraint (\ref{pdiffeo2})
\begin{align}
\delta_2 K_x=&\frac{M}{x^2}\delta_2\Kf+\delta_2\Kf'-\frac{1}{4x^2}\sqrt{\frac{2M}{x}}(x-6M)(\delta\Kf)^2 \notag \\
&+\sqrt{\frac{2M}{x}}\delta\Kf\delta\Kf'-\frac{1}{2x^2}\sqrt{\frac{2M}{x}}\mathcal{E}+\frac{4\pi }{x}p_\chi\chi'\,.  \label{delta2Kx}
\end{align}
Equation (\ref{EoMEx}) gives again
\begin{equation} \label{delta2Nx}
\delta_2 N^x=-\delta_2\Kf
\end{equation}
and  (\ref{EoMEphi}), upon using (\ref{delta2Ephi}), (\ref{delta2Kx}), (\ref{delta2Nx}) and the first-order equations,
\begin{align}
\sqrt{2Mx}\,\delta_2\dot{K}_\varphi=&-\frac{M}{x}\delta_2\Kf+2M\delta_2\Kf'-\sqrt{2Mx}\,\delta\Kf\delta\Kf' \notag \\
&+\frac{1}{x}\sqrt{\frac{2M}{x}}\mathcal{E}-\sqrt{\frac{2M}{x}}\mathcal{E}'+4\pi p_\chi\chi'+\dot{\mathcal{E}}\,. \label{delta2EphiPDE}
\end{align}

On the other hand, equation (\ref{EoMKphi}) for the time evolution of the extrinsic curvature  $\Kf$ gives
\begin{align}
\delta_2\dot{K}_\varphi=&-\frac{1}{2x}\sqrt{\frac{2M}{x}}\,\delta_2\Kf+\sqrt{\frac{2M}{x}}\,\delta_2\Kf'-\delta\Kf\delta\Kf' \notag\\
&+\frac{1}{x^2}\mathcal{E}-\frac{2\pi }{\alpha}\left(\frac{\nu}{x^3}p_\chi^2+\sigma x\chi'\,^2-xU\right) \label{delta2KphiPDE}
\end{align}
comparing each term of this equation with (\ref{delta2EphiPDE}) we
must, for consistency, have the identity
\[
\dot{\mathcal{E}}=\sqrt{\frac{2M}{x}}\mathcal{E}'-4\pi  p_\chi\chi'-2\pi \frac{\sqrt{2Mx}}{\alpha}\left(\frac{\nu}{x^3}p_\chi^2+\sigma x\chi'\,^2-xU\right)
\]
or, simplifying the RHS using (\ref{effectiveE}),
\begin{equation}\label{Edot1}
\dot{\mathcal{E}}=-4\pi \left[\left(1-\frac{2M}{x}\right)p_\chi\chi'+\frac{\sqrt{2Mx}}{\alpha}\left(\frac{\nu}{x^3}p_\chi^2+\sigma x\chi'\,^2\right)\right]+\text{surface term}\,.
\end{equation}

That this is indeed the case can be readily verified using  the (first order) equations of motion for the matter field, (\ref{EoMchi}) and (\ref{EoMpchi}) or
\begin{equation}
\dot{\chi}=\frac{\nu}{\alpha x^2}p_\chi+\sqrt{\frac{2M}{x}}\,\chi' \quad,\quad
\dot{p}_\chi=\left(\frac{\sigma x^2}{\alpha}\chi'\right)'-\frac{x^2}{2\alpha}\frac{{\rm d}U}{{\rm d}\chi}+\left(\sqrt{\frac{2M}{x}}\,p_\chi\right)' \label{matterEqs}
\end{equation}
in the present gauge, to compute the time derivative of $\mathcal{E}$
from its definition (\ref{effectiveE}):
\begin{equation} \label{Edot2}
\dot{\mathcal{E}}=4\pi \int_x^\infty {\rm d}z\left[\left(\frac{\nu\sigma}{\alpha^2}-\frac{2M}{z}\right)p_\chi\chi'+\frac{\sqrt{2Mz}}{\alpha}\left(\frac{\nu}{z^3}p_\chi^2+\sigma z\chi'\,^2\right)\right]'\,.
\end{equation}
Comparing (\ref{Edot1}) and (\ref{Edot2}), we see here how the anomaly-freedom condition (\ref{anomalyFreeCond}) is required for consistency.

Once anomaly freedom is implemented, equations of motion can be
consistently used to evaluate the dynamics even in the absence of a
classical space-time structure. We will now turn to the issue of
horizons, whose primary motivation and definition is closely tied to
classical space-time intuition.

\subsubsection{Classical horizon conditions}

We introduce inverse-triad corrections in the Hamiltonian constraint
by replacing $N/\sqrt{\ex}$ by $N\alpha/\sqrt{\ex}$.
For the Schwarzschild gauge this can be accounted for most simply by setting:
\be
N\alpha = \sqrt{1-\frac{2M}{x}}
\ee
and replacing $\rho$ by $\rho_{\rm mod}/\alpha$ where
$\rho_{\rm mod}$ contains further corrections such as $\nu$ and
$\sigma$ used above for a
scalar field. 
By following the procedure in Sec.~\ref{s:S} and simple substitution 
in \eqref{horcond} we have
\be \label{ModHorizonSchwarz}
1-\frac{2M}{x} + \frac{2}{x}\int_x^{\infty}{\dif z
  \,4\pi\frac{\rho_{\rm mod}}{\alpha} z^2} -(\alpha^2 - 1 )(\dkp)^2=0\,.
\ee
Now, $\delta K_{\varphi}$ no longer cancels because different powers
of $\alpha$ appear in the terms of (\ref{horcond}) with different
powers of $N$ in the denominators.  The isolated horizon condition
gives the results from \eqref{schwisohorcondeq1} and
\eqref{schwisohorcondeq2}, with $\rho$ replaced by $\rho_{\rm
mod}/\alpha$, and $\delta K_{\varphi}$ vanishes by definition. Thus
the two horizon conditions give different results, becoming
equivalent only in the case when $\dkp = 0$. One may choose this value
to fix the Hamiltonian gauge, but the more general condition of
trapping horizons remains gauge dependent.

For the
Painlev\'e--Gullstrand gauge, we have, again
up to second order, the horizon condition (\ref{horcond}) as
\begin{align} \label{ModHorizonPG}
1-\alpha^2\frac{2M}{x}+(\alpha^2-1)\left(2\sqrt{\frac{2M}{x}}\delta\Kf+2\sqrt{\frac{2M}{x}}\delta_2\Kf-(\delta\Kf)^2\right)+\frac{2}{x}\mathcal{E}=0\,.
\end{align}
In contrast to the Schwarzschild case with the same correction in the
Hamiltonian constraint, even the background terms are modified as a
consequence of the term $(N^x/N)^2$ in (\ref{horcond}), now with a
non-vanishing shift vector. Different slicings do not give rise
to the same area-mass relationship of horizons, further illustrating
the gauge dependence of the original horizon condition.

\subsubsection{Horizon conditions for modified space-time structures}

The case of a modified, yet consistent constraint algebra provides
several interesting lessons. Not only do the horizon conditions
we use lead to different results (\ref{ModHorizonSchwarz}) and
(\ref{ModHorizonPG}) for different choices of slicings, for each
slicing they depend on the gauge-dependent quantity $\delta
K_{\varphi}$. With this dependence, the horizon conditions are no longer
meaningful. The application of conventional space-time intuition to
quantum gravity, embodied here by some of its effects on modified
constraints, is thus highly non-trivial. In Section~\ref{s:Disc} we
will discuss this set of problems and its ramifications further.  

We recall that the modified equations are fully consistent
dynamically; it is only the horizon conditions which must be adapted
as well by using as yet unknown notions of quantum horizons. To
provide an idea of the required modifications of horizon conditions,
it turns out that the modified trapping-horizon condition
\be \label{ModHor}
\frac{\ex}{(E^\varphi)^2}-\left(\frac{N^x}{\bar{\alpha}N}\right)^2=0
\ee
when evaluated for all cases considered here produces satisfactory
results: there is no gauge dependence in the area-mass relationships, and
they all agree for the different slicings, correcting the classical
relationship by
\begin{equation}
1-\frac{2M}{x} + \frac{2}{x}\int_x^{\infty}{\dif z
  \,4\pi\frac{\rho_{\rm mod}}{\alpha} z^2}=0\,.
\end{equation}
Moreover, the corrections differ from those found in the
non-absorbable case with classical constraint algebra, where
we have (\ref{HorizonfS}). 

The combination of fields appearing in the modified horizon condition
may be interpreted as the inverse-metric component $g^{xx}$ for a
metric with rescaled lapse function $\bar{\alpha}N$, but in the case
of a modified constraint algebra the notion of line elements or
metrics is not applicable. Instead, the modification can be read off
from the dynamical equations used here, ensuring that evaluations for
horizons are gauge invariant. The isolated-horizon condition fixes the
Hamiltonian gauge before quantization or putting in corrections, and
thus removes the gauge-dependent term by fiat. This form of gauge
fixing before quantization, or before including corrections,
eliminates important consistency conditions, and thus, if it is used
as the sole means to determine horizons, further necessary conditions
to the horizon condition such as (\ref{ModHor}) would be overlooked.

\subsection{Modified algebra, non-absorbable}

The equations for $1\not=\alpha\not=\bar\alpha\not=1$ can be mapped to
those analyzed in Section \ref{s:Class} by absorbing $\bar{\alpha}$ in
the lapse function. We can thus skip analyzing this general case anew
and simply cite the conclusions drawn earlier: Corrections to the
area-mass relation do arise, even in vacuum space-times. However, as
in Section \ref{s:ModAbs}, absorbing a correction function in the
lapse function makes the horizon conditions differ in the two
definitions used here, and gauge-dependent terms no longer drop out,
unless the horizon condition is corrected to (\ref{ModHor}). Combining
the previous area-mass relationships, we arrive at
\begin{equation} \label{HorizonGeneral}
 1-\frac{2Mf_{\alpha/\bar{\alpha}}}{x} +
\frac{2f_{\alpha/\bar{\alpha}}}{x}\int_x^{\infty}{\dif z
 \;\frac{4\pi\rho}{f_{\alpha/\bar{\alpha}}\bar{\alpha}} z^2} =0
\end{equation}
where $f_{\alpha/\bar{\alpha}}$ is computed as in the case of
$\bar{\alpha}=1$, but replacing $\alpha$ with $\alpha/\bar{\alpha}$.

\section{Discussion}
\label{s:Disc}

When quantum gravity changes the structure of space and time, as
expected in many different ways at a fundamental level, the usual
notions of geometry and physical implications for instance in the
behavior of black holes must be reanalyzed. In particular, one cannot
always make use of definitions that refer directly or indirectly to
space-time manifolds or even coordinates. The line element, one of the
basic concepts often used in classical general relativity, is the main
example for this; and constructions based on its properties such as
some notions of horizons cannot always be applied in the presence of
quantum-gravity corrections. But even if one does not rely on line
elements or metric components, the concept of a horizon crucially
refers to test-particle propagation in space-time (e.g.\ for trapping
surfaces or causal properties). The notion of test particles does not
exist in fundamental quantum-gravity theories, and even at effective
levels this notion can lead to additional difficulties if space-time
structures change.\footnote{For instance, in \cite{ModCollapse}
  apparent superluminal effects arise, but only because the space-time
  notion used for null lines is not applicable for the deformed
  constraint algebra.}

In this article, we have illustrated some of these features by
different examples of inverse-triad corrections in spherically
symmetric models of loop quantum gravity, showing the various ways in
which the area-mass relationship of horizons is modified by
inverse-triad corrections. While our calculations of the dynamics are
not at the full quantum level of the theory, which is still too
difficult to handle explicitly, several features such as modified
space-time structures as evidenced by non-classical constraint
algebras, can be highlighted. This led us to stress the importance of
rethinking definitions of horizons suitable for quantum gravity.

In order to probe properties of black-hole horizons in a more
general context, allowing for corrections to the constraint algebra, we
have developed a canonical version of spherically symmetric
perturbation theory in connection variables. Several perturbation
equations can be solved completely in the presence of matter,
providing general formulas for the dynamics of trapping horizons. In
the classical case, these formulas are not new, but their new
derivation allows an easy extension to geometries arising from
canonical quantizations and the related modified space-time
structures.

Quantum-gravity corrections, from this perspective, can be split into
two classes: those that modify the dynamics of general relativity but
not its space-time structure, leaving the classical constraint algebra
unchanged; and those that modify both the dynamics and the space-time
structure. We have presented a detailed analysis of a model falling in
the former class, where a standard space-time analysis is available in
the presence of inverse-triad corrections, used for the results
presented in Sec.~\ref{s:Undeformed}. As seen there, the horizon
behavior is affected by the corrections, for instance regarding the
relationship between mass and size, or Hawking radiation. But the
classical notion of a horizon is still valid, illustrated by the
result of Section \ref{s:Class} that different horizon conditions
agree with each other and are gauge independent. Moreover, in this
case ($\bar{\alpha}=1$) the canonical horizon conditions produce the
same result as a direct space-time analysis.

We have not attempted to address the question of how in general to
define horizons in modified space-time, but we have provided an
example where direct extensions of classical conditions fail when
quantum gravity modifies space-time structures. Properties of horizons
according to the classical definitions then depend on the slicing
chosen, and are gauge dependent. In the examples considered here, a
simple modification of the classical horizon conditions (\ref{ModHor})
by the correction function that also changes the dynamics leads to
satisfactory results. In particular, the area-mass relationship
  is corrected to the implicit condition
\begin{equation}
 R_{\rm hor}= 2f_{\alpha/\bar{\alpha}}(R_{\rm hor})\left(M-
 \int_{R_{\rm hor}}^{\infty} {\rm d}x 4\pi x^2
 \frac{\rho(x)}{f_{\alpha/\bar{\alpha}}(x) \bar{\alpha}(x)}\right)
\end{equation}
for the area radius $R_{\rm hor}$ of the horizon, with
$f_{\alpha/\bar{\alpha}}$ related to the primary correction functions
$\alpha$ and $\bar{\alpha}$ by
$f_{\alpha/\bar{\alpha}}'/f_{\alpha/\bar{\alpha}}=
(1-\alpha/\bar{\alpha})/x$. 

No gauge-dependence appears in the condition for the horizon radius,
and the different slicings lead to equivalent results. But the
modified horizon condition was not obtained by quantum space-time
intuition; rather, we looked for a modification that served to
eliminate gauge-dependent terms. Our results especially in the case of
modified yet first-class constraint algebras, the general case
expected for loop quantum gravity, thus show the need to develop
appropriate horizon definitions for quantum space-times without
referring to the usual classical notions such as the expansion of
light rays which are no longer available. Some steps in this direction
have already been undertaken, for instance in
\cite{HWHorizon,TriangBH,BHCohHor,TimeHor} and recently in
\cite{LQGHor}, but most of them remain tied to the classical notion of
expansion and they are difficult to evaluate in a dynamical
context. Our results also show that the more restrictive notion of
isolated horizons, based on an additional gauge fixing compared with
trapping horizons, does not seem sufficient to derive corrected
horizon conditions.

Our considerations provide a cautious note regarding the reliability
of black-hole entropy calculations in loop quantum gravity, which are
based on a classical implementation of isolated horizons treated as
boundaries of space-time \cite{ABCK:LoopEntro}.  The properties of
horizon definitions found here indicate that the implementation of
isolated horizons via boundary conditions derived before quantization
may not include all possible quantum features relevant for
horizons. Even though quantum-gravity corrections are expected to be
small for realistic black holes, the value of the Barbero--Immirzi
parameter derived from entropy countings could change. In particular,
it is not clear whether a universal value of the parameter,
independent of the type of black hole, would still arise. In this way,
new interesting and non-trivial tests of the quantization may be
possible. On the other hand, as a supportive statement for some of the
assumptions behind the current counting procedures, our results for
the case of quantum effects leaving the classical constraint algebra
intact also show that corrections to the area and temperature laws
arise from modifications in the dynamics even if classically motivated
horizon conditions are used. The fact that, at least in some cases,
classical definitions can consistently be used even for the
quantum-modified dynamics shows, among other things, that a possible
renormalization of Newton's constant, as sometimes suggested
\cite{HorizonRenorm}, need not necessarily be taken into account for
the horizon condition itself (or for countings of entropy based on
it);\footnote{There may be other motivations to introduce
renormalization at the level of horizon conditions independent of the
present context.}  it will in any case arise once horizon conditions
are evaluated for a dynamical solution, producing the area-mass
relationship. Inverse-triad corrections, considered here as an
important contribution from quantum geometry, do not constitute the
usual source of renormalization. But the canonical methods developed
and applied here can also be used for quantum back-reaction, which in
its canonical form formulated in \cite{EffAc,Karpacz} corresponds to
the familiar quantum-dynamical corrections of interacting quantum
theories. Our results thus provide a first step toward possible
implications of renormalization in dynamical solutions of loop quantum
gravity.

\begin{appendix}

\section{Space-time transformations with modified constraint algebra}

In this appendix, we compare different coordinate representations of
solutions in the case of a modified constraint algebra, showing that
they are not related by coordinate transformations. To be specific, we
choose the absorbable case $\alpha=\bar{\alpha}\not=1$.

\subsection{Schwarzschild-like}

A Schwarzschild-like solution can be obtained by assuming
$K_{\varphi}=K_{x}=N^{x}=0$. Since the vacuum Hamiltonian-constraint
equation is the same as in the classical case we have the
Schwarzschild solution for $E^{\varphi}$ if we assume the gauge
$E^{x}=x^{2}$. Only the form of the lapse function changes and using
\eqref{EoMKphi} is found to be $N=\alpha^{-1}(1-2M/x)^{1/2}$, as
already suggested by the absorbable nature of the inverse-triad
correction in the case under consideration. If we were to assume
that even with the modified algebra there is a spacetime interpretation,
we would write the solution as the corresponding Schwarzschild-like
line element
\begin{equation}\label{Schwarz}
\not\!{{\rm d}}s^{2}=-\alpha^{-2}\left(1-\frac{2M}{x}\right)\not\!{\rm d}t^{2}+
\left(1-\frac{2M}{x}\right)^{-1}\not\!{\rm d}x^{2}+x^{2}\not\!{\rm
  d}\Omega^{2} \,. 
\end{equation}
(The slashed ds indicate that the line element in the present
  context is a purely formal construction, with $\not\!{\rm d}x^a$ not
  subject to the usual coordinate transformations.)

\subsection{Painlev\'e--Gullstrand-like}

Following the analysis of section 3.1.2 we now consider the
transformation to a Painlev\'e--Gullstrand like metric. Since
(\ref{Schwarz}) is time independent, there is a timelike Killing vector
$\xi_{(t)}=\partial_{t}$. If $u^a$ is the tangent to a radial
freely falling geodesic (parameterized by $T$) then
$u_a\xi^a_{(t)}=E$, where $E$ is a constant which we
choose to be equal to one. This implies
\[
g_{ab}u^a\xi^b_{(t)}=
-\alpha^{-2}\left(1-\frac{2M}{x}\right)\frac{{\rm d}t}{{\rm d}T}=-1
\]
or, with $g_{ab}u^au^b=-1$,
\[
\frac{{\rm d}x}{{\rm d}T}=-\sqrt{\alpha^{2}-1+\frac{2M}{x}}\,.
\]
The time differential ${\rm d}T=-u_adx^a$ with
$u_a=(-1,-(1-2M/x)^{-1}\sqrt{\alpha^{2}-1+2M/x},0,0)$ reads
\[
{\rm d}t={\rm
  d}T-\left(1-\frac{2M}{x}\right)^{-1}\sqrt{\alpha^{2}-1+\frac{2M}{x}}
{\rm d}x\,.
\]
Substituting this back in the Schwarzschild metric we obtain
\begin{equation} \label{PGds}
\not\!{\rm d}s^{2}=-\alpha^{-2}\left(1-\frac{2M}{x}\right)\not\!{\rm d}T^{2}+
\alpha^{-2}\not\!{\rm d}x^{2}+2\alpha^{-2}\sqrt{\alpha^{2}-1+\frac{2M}{x}}
\not\!{\rm d}x\not\!{\rm d}T+x^{2}\not\!{\rm d}\Omega^{2}
\end{equation}
which can be considered as the Painlev\'{e}--Gullstrand version of
(\ref{Schwarz}). 

For our phase-space functions, (\ref{PGds}) implies $E^{x}=x^{2}$,
$E^{\phi}=x/\alpha$, $N^{x}=\sqrt{\alpha^{2}-1+2M/x}$, $N=1$,
$K_{\phi}=-\sqrt{\alpha^{2}-1+2M/x}/\alpha$,
$K_{x}=(2M\alpha'+M\alpha/x-\alpha'x)/\alpha^{3}x\sqrt{\alpha^{2}
-1+2M/x}$. However, substituting this form of the metric back in the
constraints we find that the diffeomorphism constraint satisfied, but
not the Hamiltonian constraint. This is an illustration of the fact
that the modified form of the constraint algebra prevents coordinate
transformations from being gauge transformations: they do not map
solutions of the constraints to other solutions.  With a version of
inverse-traid corrections not modifying the constraint algebra, on the
other hand, the analysis of Section \ref{s:ModPG} showed that the
metric in the new coordinates did satisfy all constraints and was a
solution representing the same spacetime.

Earlier, we have seen that $N\alpha=1$ solves the Hamiltonian
constraint, but it does not correspond to the Painlev\'e--Gullstrand
form obtained by following the spacetime procedure to transform from
the Schwarzschild metric.
As discussed in Section \ref{constraints}, absorbing the correction
function in the lapse function does not amount to reducing the
constraint algebra to classical form. Conversely to the transformation
attempted here, one may start with the Painlev\'{e}--Gullstrand-like
solution solving the constraints and transform to some Schwarzschild
form. For the static form of the Schwarzschild line elements combined
with our usual gauge fixing of $E^x$, two coefficients, $g_{tt}$ and
$g_{xx}$, have to be determined. If the Painlev\'{e}--Gullstrand form
is given, one may follow the procedure used above backwards, asking
what Schwarzschild-like coefficient would provide the desired
Painlev\'{e}--Gullstrand form in this way. With three non-trivial
coefficients to be produced for the Painlev\'{e}--Gullstrand form, but
only two free coefficients for a Schwarzschild-like form, three
equations for two unknowns must be solved. Classically, there is a
consistent solution, but there is none when the constraints of a
modified algebra are used.

\end{appendix}

\section*{Acknowledgements}

RT thanks Ghanshyam Date for useful discussions. This work was supported in part by NSF grant 0748336.


\newcommand{\noopsort}[1]{}

\end{document}